\documentclass[11pt]{article}

\usepackage[american]{babel}
\usepackage{csquotes}
\usepackage{graphicx}
\usepackage{amsmath}
\usepackage{amsfonts}
\usepackage{float}
\usepackage{hyperref}
\usepackage{enumitem}
\usepackage{bbm}
\usepackage{tikz}
\usetikzlibrary{positioning,calc}
\usepackage{adjustbox}
\usepackage[
  style=apa
]{biblatex}

\newcommand{\bs}{\boldsymbol}
\newcommand{\bPsi}{\boldsymbol{\Psi}}

 \newenvironment{psmallmatrix}
  {\left(\begin{smallmatrix}}
  {\end{smallmatrix}\right)}

\begin{document}


\thispagestyle{empty} 
\baselineskip=28pt

\begin{center}
{\LARGE{\bf  Mixtures of Neural Network Experts with Application to Phytoplankton Flow Cytometry Data}}

\end{center}

\baselineskip=12pt

\vskip 2mm
\begin{center}


Ethan Pawl\footnote{(\baselineskip=10pt to whom correspondence should be
  addressed) Department of Statistics, University of California, Santa Cruz, 1156 High St.,
  Santa Cruz, CA 95064, epawl@ucsc.edu}
Fran\c{c}ois Ribalet\footnote{\baselineskip=10pt School of Oceanography, University of Washington, Seattle, 1410 NE Campus Pkwy.,
  Seattle, WA 98195,
  ribalet@uw.edu}
Paul A. Parker\footnote{\baselineskip=10pt Department of Statistics, University of California, Santa Cruz, 1156 High St.,
  Santa Cruz, CA 95064, paulparker@ucsc.edu}\, and
Sangwon Hyun\footnote{\baselineskip=10pt Department of Statistics, University of California, Santa Cruz, 1156 High St.,
  Santa Cruz, CA 95064,
  sangwonh@ucsc.edu}

\end{center}

\vskip 4mm
\baselineskip=12pt 
\begin{center}
{\bf Abstract}
\end{center}

\noindent
Flow cytometry is a valuable technique that measures the optical properties of particles at a single-cell resolution. When deployed in the ocean, flow cytometry allows oceanographers to study different types of photosynthetic microbes called phytoplankton. It is of great interest to study how phytoplankton properties change in response to environmental conditions. In our work, we develop a nonlinear mixture of experts model to estimate separate regression functions for each subpopulation utilizing random-weight neural networks. Our model allows one to flexibly estimate how cell properties and relative abundances depend on environmental covariates in each segment of a heterogeneous sample, without the computational burden of backpropagation. We show that the proposed model provides superior predictive performance in simulated examples compared to a mixture of linear experts. Also, applying our model to real data, we show that our model has (1) comparable out-of-sample prediction performance, and (2) more realistic estimates of phytoplankton behavior.

\baselineskip=12pt
\par\vfill\noindent
\par\medskip\noindent

\clearpage\pagebreak\newpage \pagenumbering{arabic}
\baselineskip=24pt

\begin{refsection}

\section{Introduction}

Phytoplankton are photosynthetic microbes that reside at the bottom of the marine food chain. They conduct more photosynthesis than all flora on land \parencite{field1998pp}, acting as an invaluable component of Earth's biological carbon pump \parencite{Ducklow2015,Boyd2019}. Thus, studying phytoplankton behavior and distribution in the ocean is directly relevant for our understanding of carbon transport and, by extension, the earth's biogeochemical cycle \parencite{phyto_pump,phyto_pump_2}. Phytoplankton have faced a number of environmental pressures in recent years, including increased levels of carbon dioxide, subsequent ocean acidification, and higher average annual temperatures \parencite{van2020stressors}. Some recent work predicts major changes in relative abundances \parencite{flombaum2021diverse} in response to environmental changes. However, it is still largely unknown how these trends will affect phytoplankton abundance and photosynthetic activity. Thus, understanding the response of phytoplankton to various environmental conditions constitutes an important line of work.

An increasingly popular method to measure phytoplankton attributes in the ocean is \textit{flow cytometry} \parencite{sosik2003cyto}.  A fluid sample is passed through an instrument which measures the size and fluorescence for each particle. Then, the data analyst uses expert judgment -- aided by software -- to group these particles' measurements into distinct types. This process is called \textit{gating} and has traditionally been done manually, heavily relying on the analyst's visual inspection of the data \parencite{Verschoor2015Jul, flowjo}. In recent years, automatic gating algorithms have been in active development as an alternative to manual gating. However, most currently available gating algorithms are motivated by biomedical applications and are unsuitable for marine flow cytometry data. Our primary focus is to quantify each species' response to changes in the ocean's environmental conditions in a flexible manner, using a custom statistical mixture model. We adopt the mixture of experts framework to separate particles into subpopulations, fit a separate regression on each subpopulation, and also fit a regression on each cluster probability. This allows a simple framework for domain experts to understand the relationships between all subpopulation properties and environmental conditions.

We briefly review the literature on statistical models for flow cytometry that are most relevant to our work. Recent \textit{model-based} statistical methods typically assume the data come from a mixture of real-valued statistical distributions. \textcite{Pyne2014Jul} and \textcite{Lee2016Jan} use finite mixtures of multivariate skew $t$-distributions with sample-specific random effects. \textcite{Dundar2014Dec} and \textcite{Cron2013} use hierarchical Dirichlet process mixtures of Gaussian kernels to borrow information across samples. \textcite{Soriano2019Mar} develop a specialized Dirichlet process Gaussian mixture which accounts for inter-sample variation in both the component parameters and weights. \textcite{Gorsky2024Coarsened} develop a Bayesian nonparametric mixture of hierarchical skew normal kernels, a hierarchy which captures both sample-specific and inter-sample structure. \textcite{Hejblum2019Mar} develop a Dirichlet process mixture of skew $t$-distributions which accounts for temporal ordering of samples. Specific to oceanography, \textcite{Hyun2025} use an expectation-maximization (EM) algorithm to estimate a trend-filtered mixture of Gaussians which automatically gates cells into subpopulations. \textcite{deSousa2025} build on this technique by providing a more computationally lightweight solution via kernel smoothing. However, none of these approaches use environmental covariates. Thus, these models are suitable for clustering a population into subpopulations, not for inference on the response of each subpopulation to environmental changes, which is our primary interest. The most relevant method, outlined in \textcite{flowmix}, applies a mixture of experts \parencite{Jordan1994Mar} to phytoplankton flow cytometry data, modeling both cluster means and probabilities as linear functions of environmental covariates. We build on this last method by developing a mixture of \textit{nonlinear} experts.

The use of mixtures of experts to jointly perform regression and clustering is supported by a rich literature which has developed over the past thirty years.  Recently, \textcite{Huang2012Jun} showed that a semiparametric mixture of experts is identifiable under weak assumptions, and \textcite{Nguyen2021Dec} showed that a mixture of experts using a softmax gating function and location-scale kernels can provide arbitrarily accurate density estimates. Also, the use of a neural network in a mixture of experts has gained popularity in recent years. For example, \textcite{Hadavandi2016May} use a mixture of neural network experts in an agglomerative clustering scheme, and \textcite{Liu2020} propose a mixture of neural network regressions in which all neural network experts share a common set of parameters. However, the latter approach does not use covariates to inform the cluster probabilities, and both are computationally intensive to train, requiring algorithms such as backpropagation or batch-size gradient descent. Our model also uses a mixture of neural network experts, but we randomly draw and then fix the hidden layer weights, thus bypassing the need for heavy computation. \textcite{ELMFunctions} show that such random-weight neural networks can approximate any continuous data-generating function with arbitrary accuracy and still generalize well to held-out samples. This technique has been applied to a variety of settings, including nonlinear regression \parencite{He2016Oct}, clustering \parencite{He2014Mar,Ding2017Apr}, and mixture modeling \parencite{Husmeier1998Jan}. See \textcite{Cao2018Jan} for a more thorough review.

Although mixtures of experts provide a natural framework for covariate-driven clustering, related latent variable models have also been developed for temporal and spatial data. Switching regression models associate each observation with a latent state corresponding to distinct regression functions, where the latent state evolves according to an unobserved Markov process \parencite{Goldfeld1973}. In longitudinal settings, discriminant-based methods have been developed to cluster \parencite{Kakizawa1998} or classify \parencite{Huang2004} a set of multivariate time series based on similarities between their spectral densities. More recently, mixtures of experts have been adapted to incorporate autoregressive time series \parencite{FS2008,Carvalho2005a,Carvalho2005b,Zeevi1996}, repeated measurements \parencite{Muthen1999}, and dynamic linear models \parencite{Munezero2023}, which allow the latent structure to evolve with respect to time, covariates, or both. \textcite{Nguyen2016} extend similar ideas to the spatio-temporal setting by combining mixtures of autoregressions with a Markov random field to induce spatial clustering. However, these approaches generally assume observations within a time series correspond to measurements of the same entity across time, which may transition between states at different times. In contrast, flow cytometry data consist of distinct biological samples collected at different times and locations with varying environmental conditions. In other words, no two measurements in a flow cytometry dataset can be matched to the same cell, as illustrated in Figure \ref{fig:distReg}. Consequently, rather than recovering the trajectory of an individual organism or population through latent states, our goal is to infer how the distribution of each latent subpopulation's biological traits changes as a function of environmental covariates. This perspective is closely related to distributional regression, in which covariates are used to model changes in features of the response variable's distribution, including location, scale, and shape \parencite{Rigby2005}. Mixtures of experts provide a flexible extension of this idea by allowing covariates to influence both the mixing probabilities of latent subpopulations and the parameters governing each subpopulation's distribution \parencite{Rugamer2024}.

\begin{figure}
    \centering
    \begin{adjustbox}{max width = \linewidth}
    \begin{tikzpicture}[
        >=latex,
        font=\small,
        sample/.style={
            draw,
            circle,
            minimum size=2.8cm,
            align=center
        }
    ]
    
    \def\gauss#1#2#3{#1*exp(-((\x-#2)^2)/(2*#3^2))}
    
    
    \begin{scope}[shift={(0,0)}]
    
    \node at (2,-0.7) {Time $t-1$};
    
    \draw[->] (0,0)--(4.3,0);
    \draw[->] (0,0)--(0,1.7);

    \node[font=\scriptsize] at (0,2)
    {$f\left(y_\cdot^{(t-1)}\mid x_{t-1}\right)$};

    \node[font=\scriptsize] at (4.85,0.075)
    {$y_\cdot^{(t-1)}$};
    
    
    \draw[very thick]
    plot[smooth,domain=0:4,samples=150]
    (\x,{0.95*(\gauss{1.20}{0.60}{0.35}
              +\gauss{0.45}{3.60}{0.40})});
    
    
    \node[sample] (S1) at (2,3.75)
    {
    $\begin{array}{c}
    y^{(t-1)}_{1}\\
    \vdots \\
    y^{(t-1)}_{n_{t-1}}
    \end{array}$
    };
    
    \draw[->] (2,1.2)--(S1);
    \node[font=\scriptsize] at (3.15, 1.775) 
    {$n_{t-1}$ \textit{iid} draws};
    
    \end{scope}
    
    
    \draw[->] (4.6,0.85)--(5.6,0.85);
    
    
    \begin{scope}[shift={(6,0)}]
    
    \node at (2,-0.7) {Time $t$};
    
    \draw[->] (0,0)--(4.3,0);
    \draw[->] (0,0)--(0,1.7);

    \node[font=\scriptsize] at (0,2)
    {$f\left(y_\cdot^{(t)}\mid x_t\right)$};

    \node[font=\scriptsize] at (4.85,0.075)
    {$y_\cdot^{(t)}$};
    
    \draw[very thick]
    plot[smooth,domain=0:4,samples=150]
    (\x,{0.95*(\gauss{0.80}{1.00}{0.35}
              +\gauss{0.80}{3.00}{0.35})});
    
    
    \node[sample] (S2) at (2,3.75)
    {
    $\begin{array}{c}
    y^{(t)}_{1}\\
    \vdots \\ 
    y^{(t)}_{n_t}
    \end{array}$
    };
    
    \draw[->] (2,1.2)--(S2);
    \node[font=\scriptsize] at (3, 1.775) 
    {$n_{t}$ \textit{iid} draws};
    
    \end{scope}
    
    
    \draw[->] (10.6,0.85)--(11.6,0.85);
    
    
    \begin{scope}[shift={(12,0)}]
    
    \node at (2,-0.7) {Time $t+1$};
    
    \draw[->] (0,0)--(4.3,0);
    \draw[->] (0,0)--(0,1.7);

    \node[font=\scriptsize] at (0,2)
    {$f\left(y_\cdot^{(t+1)}\mid x_{t+1}\right)$};

    \node[font=\scriptsize] at (4.85,0.075)
    {$y_\cdot^{(t+1)}$};
    
    \draw[very thick]
    plot[smooth,domain=0:4,samples=150]
    (\x,{0.95*(\gauss{0.55}{1.60}{0.40}
              +\gauss{1.2}{2.60}{0.30})});
    
    
    \node[sample] (S3) at (2,3.75)
    {
    $\begin{array}{c}
    y^{(t+1)}_{1}\\
    \vdots \\
    y^{(t+1)}_{n_{t+1}}
    \end{array}$
    };
    
    \draw[->] (2,1.2)--(S3);
    \node[font=\scriptsize] at (3.15, 1.775) 
    {$n_{t+1}$ \textit{iid} draws};
    
    \end{scope}
    
    \end{tikzpicture}
    \end{adjustbox}
    \caption{\textit{Illustration of the assumed data-generating process. We assume the data arise from a mixture of Gaussian components, with component means and mixing probabilities evolving according to the values of the covariates $x_t$. The lower panels depict the underlying covariate-dependent response distribution across time. As a basic example, we display a two-component mixture whose component means move closer together over time, while the leftmost component's mixing probability decreases. At each time $t$, $\{y_1^{(t)}, \cdots, y_{n_t}^{(t)}\}$ is an independent sample from the covariate-dependent response distribution, $f(y_\cdot^{(t)} \mid x_t)$. Since the research cruise is at different locations at times $t$ and $t+1$, we assume different particles are sampled at each time (e.g., $y_1^{(t)}$ and $y_1^{(t+1)}$ are not measurements of the same particle).}}
    \label{fig:distReg}
\end{figure}

The method of \textcite{flowmix} is the first such approach developed specifically for oceanographic flow cytometry. It assumes Gaussian cluster means and the logits of the cluster probabilities are linear functions of the covariates. However, we show that the nonlinear functions are more apt for modeling flow cytometry data collected over a large spatial region. For instance, microbe characteristics such as cell size have biological upper and lower limits 
which linear functions cannot predict effectively. Generally, phytoplankton species' attributes and abundance likely do not respond to environmental changes monotonically.
Also, certain pairs of covariates may have an interaction effect -- for example, increased nutrient availability is known to improve or hinder a cell's ability to perform biological functions. 
Apart from these general hypotheses, very little is known about the functional forms of these nonlinear relationships. This lack of prior knowledge makes our flexible nonparametric approach to the regression functions useful.

Here, we extend the approach of \textcite{flowmix} in two ways. First, we use a random-weight neural network to expand the class of estimable regression functions. Such a model can more flexibly model cluster means and probabilities as functions of environmental conditions, thus allowing the user to fit a single model to flow cytometry data collected from a wide, heterogeneous range of ocean conditions. Second, we use the principal components of environmental covariates as regressors rather than the original covariates themselves. This vastly reduces the number of regressors and removes the strong intercorrelations in the original covariates, thus easing model interpretation. 

The rest of this article is organized as follows: in Section \ref{sec:data} we describe the data. In Section \ref{sec:method} we describe the motivation behind the model and the structure of the model itself. In Section \ref{sec:sim} we provide a simulation study which provides strong evidence towards the idea that the proposed model fits linear data as well as or better than a mixture of linear experts and fits strongly nonlinear data better than the linear model. In Section \ref{sec:app} we apply the proposed model to real flow cytometry data and uncover new insights into the responses of phytoplankton to changes in environmental conditions. Finally, in Section \ref{sec:disc} we review our methods and results before outlining recommendations for future work.

\section{SeaFlow cruise data description} \label{sec:data}

The flow cytometry data we analyze are from the Gradients 2 cruise collected by the Armbrust Lab \parencite{Ribalet2019Nov} in the summer of 2017. This cruise traversed a large spatial region with heterogeneous environmental conditions, making it a good candidate to showcase the flexibility of our model to variation in the relationships between response data and covariates. Specifically, the cruise traveled between the North Pacific Subpolar and Subtropical gyres, collecting measurements of phytoplankton abundance and optical properties using the continuous-time flow cytometer known as SeaFlow \parencite{seaflow}. We refer to each observation collected by SeaFlow as a \textit{particle}, since it is a collection of optical properties of a particle suspended in ocean water. We assume particles are aggregated to discrete time, so each time point $t$ contains multiple ($n_t$) particles collected over a wider time interval (in our case, an hour), for $t = 1, \ldots, T$. We denote the $i$th particle at time $t$ as $\boldsymbol{y}_i^{(t)} \in \mathbb{R}^d$, for $i = 1, \ldots, n_t$ and refer to the collection of all particles at time $t$ as a \textit{cytogram}, denoted by $\bs{y}^{(t)}$. SeaFlow measures $d = 3$ properties of each particle: (1) cell diameter, and (2) red fluorescence and (3) orange fluorescence, corresponding to the light intensities of emission spectra associated with chlorophyll and phycoerythrin pigments.  Since all three dimensions' original measurements are highly skewed, they are transformed using the natural logarithm, resulting in roughly Gaussian cluster shapes. We then linearly rescale the data to be roughly between 0 and 8 in each dimension. Imposing a common scale is useful for visualization and for specifying direction-invariant constraints on cluster movement (see \eqref{eqn:maxdev}).

In total, there are $308$ 3-dimensional cytograms, each consisting of all particles collected during one hour of the cruise. However, we later created a twelve-hour lagged version of a covariate, which led us to remove the first 12 cytograms for a final count of $T = 296$. Additionally, some observations were censored at SeaFlow's measurement limits; these points were simply removed prior to modeling. Lastly, an auxiliary measurement $\{c_i^{(t)}\}_{i,t}$, the inferred biomass of each particle, will be used. We use this cytogram and biomass data as a regression response.

Accompanying the response data, there are $33$ environmental covariates sourced from the Simons CMAP database \parencite{Ashkezari2021}, accessed via the \texttt{cmap4r} R package \parencite{cmap4r2025Mar}, and aggregated to match the hourly time resolution of the response data. 
These environmental covariate data include biochemical and physical attributes collected from satellites (and processed as long-term averages) and from on-board instruments, for a total of $p = 37$ covariates over $T = 296$ time points corresponding to unique hours. The vector of all $37$ covariate values at time $t$ is written as $\boldsymbol{x}_t$, and $\boldsymbol{X} \in \mathbb{R}^{296 \times 37}$ is the covariate matrix with row $t$ equal to $\boldsymbol{x}_t$.  For more details regarding data curation, see \textcite{flowmix}.

\begin{figure}
    \centering
    \includegraphics[width=\linewidth]{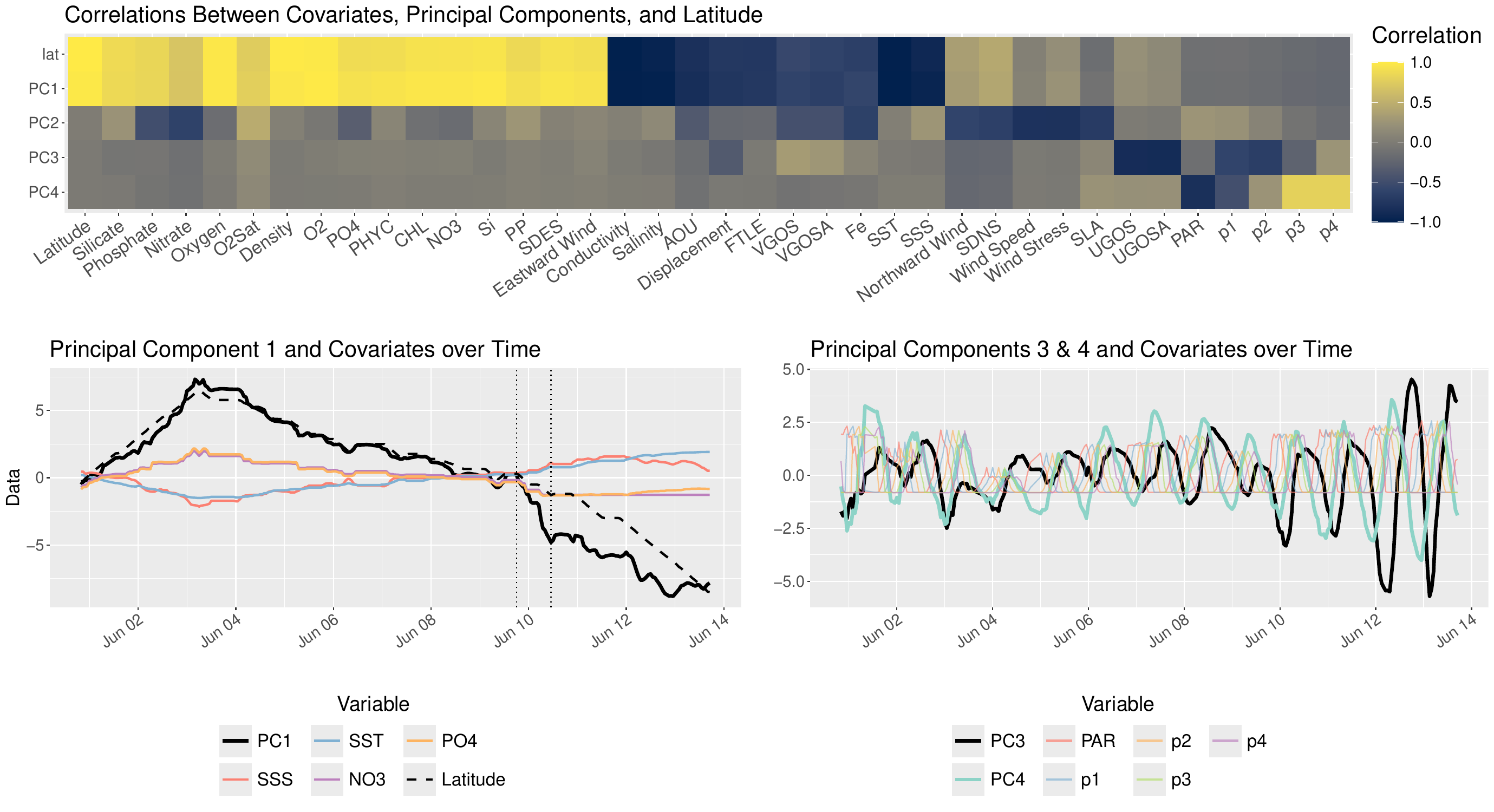}
    \caption{\textit{Principal components of the environmental covariates. The top panel shows the sample correlations between the first four principal components, latitude, and the covariates. The bottom-left panel shows the first principal component (PC1), as well as the most scientifically relevant covariates which are strongly correlated with PC1, plotted over time. We also overlay latitude as a dashed black line to show its similarity to PC1. The two vertical dotted lines at latitudes 31.9$^\circ$ N and 34$^\circ$ N represent the estimated boundaries of the North Pacific Transition Zone, a region where latitude-varying conditions change rapidly, hand-picked as the region where PC1 changes rapidly. In the bottom-right panel, we display a similar plot for PC3 and PC4. In Supplement A, we display a similar plot for PC2 (Figure S1) and the full set of principal components (Figure S2).}
    }
    \label{fig:pc_cov_corr}
\end{figure}

The 37 covariates are numerous and strongly intercorrelated, so we apply principal components analysis (PCA) to the covariate matrix. The resulting design matrix has columns consisting of the first $q$ principal components and is denoted by $\bs{\Psi} \in \mathbb{R}^{T\times q}$. In all experiments, we choose $q$ to be the lowest value such that the proportion of variance explained by the first $q$ principal components is at least 0.95. For the dataset used in our analysis, we found $q = 9$, which we use throughout this article. We show in Supplement F that model interpretation does not change much for values of $q$ larger than 9.

It is worth examining the principal components $\{\boldsymbol{\Psi}_{\cdot,j}\}_{j= 1}^4$ in some detail. For brevity we will adopt the shorthand of, for example, PC1 to denote the first principal component. As shown in Figure~\ref{fig:pc_cov_corr}, the first principal component PC1 has a strong correlation with latitude.  Since PC1 and latitude share strong associations with environmental conditions such as nutrient levels, temperature, and salinity, information about these conditions can be compressed into a single variable which summarizes latitudinal variation in the ocean. The second principal component PC2 is most strongly associated with wind speed, sea level anomaly, and iron levels, summarizing information relating to physical oceanographic characteristics. PC3 is related to physical ocean characteristics such as zonal geostrophic velocities and lagged sunlight (photosynthetically active radiation or PAR for short) time-lagged by 3 and 6 hours. Lastly, PC4 is comprised of PAR and its 9- and 12- hour lagged variants. Visual inspection of PCs 5--9 revealed no easily discernible biological interpretations, so we omit them from subsequent analysis.

\section{Methodology} \label{sec:method}

Although the cytograms $\left\{\bs{y}^{(t)}\right\}_{t = 1}^T$ are collected sequentially, it is reasonable from an oceanographic standpoint to assume that spatio-temporal variation in microbial communities is largely due to heterogeneity in environmental conditions. In comparison, variation due to time and space alone is negligible. Therefore, we assume the cytograms are conditionally exchangeable given the covariates. That is, for any permutation $\tau$ of $\{1, \cdots, T\}$, the probabilistic model for the particles is of the form
\begin{equation*}
    p\left(\bs{y}^{(\tau(1))}, \cdots, \bs{y}^{(\tau(T))} \mid \bs{x}_{\tau(1)}, \ldots, \bs{x}_{\tau(T)} \right) = p\left(\bs{y}^{(1)}, \cdots, \bs{y}^{(T)} \mid \bs{x}_1, \ldots, \bs{x}_T \right),
\end{equation*}
where $p$ is the joint density of the cytograms induced by a probabilistic model for the particles. In our mixture of experts, we assume each cytogram follows a mixture of multivariate Gaussian distributions
\begin{equation*}
    f\left(\boldsymbol{y}_i^{(t)}\right) = \sum_{k = 1}^K \pi_{k, t} \phi_d\left(\boldsymbol{y}_i^{(t)} \mid \boldsymbol{\mu}_{k, t}, \boldsymbol{\Sigma}_k\right),
\end{equation*}
where $\phi_d(\ \cdot\ | \boldsymbol{\mu}, \boldsymbol{\Sigma})$ is the probability density function of a $d$-dimensional Gaussian random vector with mean $\boldsymbol{\mu}$ and covariance matrix $\boldsymbol{\Sigma}$. Introducing a latent membership $z_i^{(t)} \in \{1,\cdots, K\}$, we can write the distribution of particle $\boldsymbol{y}_i^{(t)}$ given latent membership $z_i^{(t)}$ as:
\begin{align*}
    \left(\boldsymbol{y}_i^{(t)} | z_i^{(t)} = k \right) & \sim \mathcal{N}_d\left(\boldsymbol{\mu}_{k,t}, \boldsymbol{\Sigma}_{k}\right) \\ 
    z_i^{(t)} & \sim \text{Categorical}\left(\pi_{1,t}, \ldots, \pi_{K,t}\right),
\end{align*}
with observations independent across all particles $i$ and all times $t$. Next, we model the mixture component means and probabilities as a regression on environmental covariates. However, instead of directly regressing on the principal components of the covariates, we regress on the hidden layer outputs of a random-weight neural network \parencite{ELM}:
\begin{equation*}
\begin{aligned}
    \boldsymbol{\mu}_{k,t} & = \boldsymbol{\beta}_{0,k} + {\boldsymbol{\beta}_{k}}' \sigma\left(\boldsymbol{h}_{t}\right) \\ 
    \pi_{k,t} & = \frac{\exp\left(\alpha_{0,k} + \sigma\left(\boldsymbol{h}_{t}\right)' \boldsymbol{\alpha}_{k}\right)}{\sum_{\ell = 1}^K \exp\left(\alpha_{0,\ell} + \sigma\left(\boldsymbol{h}_{t}\right)' \boldsymbol{\alpha}_{\ell}\right)}.
\end{aligned}
\end{equation*}
Here, $\boldsymbol{\beta}_{0,k} \in \mathbb{R}^{d \times 1}$, $\boldsymbol{\beta}_k \in \mathbb{R}^{n_h \times d}$, $\alpha_{0,k} \in \mathbb{R}^{1}$, and $\boldsymbol\alpha_k \in \mathbb{R}^{n_h}$ are regression coefficients.  Also, $\sigma\left(\boldsymbol{h}_{t}\right)$ is a random-weight neural network with $n_h$ hidden nodes, constructed as follows. We draw the $m$th hidden node's weights and biases $w_{m, 1}, \cdots, w_{m, p}, b_m \overset{\text{iid}}{\sim} \text{Unif}(-a, a)$, for a fixed constant $a$. This is repeated for $m = 1, \cdots, n_h$, where $n_h$ is the total number of hidden nodes. The distribution used to generate the weights and biases (collectively called \textit{hidden weights}) should not be interpreted as a prior distribution, since the hidden weights are not updated at all during the model training process. After they are drawn, they are considered fixed, known constants. Then, the principal components are transformed as
\begin{equation} \label{eqn:hidden}
    h_{t,m} = b_m + \sum_{j = 1}^p w_{m, j} \psi_{t,j}.
\end{equation}
That is, $h_{t, m}$ is a random linear combination of the principal components at time $t$. Next, we apply a nonlinear activation function to extract nonlinear features. Specifically, collecting the hidden nodes into a vector $\bs{h}_t = (h_{t, 1}, h_{t, 2}, \ldots, h_{t, n_h})'$, we apply the logistic function element-wise: 
\begin{equation} \label{eqn:activation}
    \sigma(\boldsymbol{h}_t) = \left(\frac{1}{1 + \exp(-h_{t, 1})}, \frac{1}{1 + \exp(-h_{t, 2})},\ldots, \frac{1}{1 + \exp(-h_{t, n_h})}\right)'.
\end{equation}
The vector $\bs{h}_t$ can also be represented more succinctly as 
\begin{equation*}
    \bs{h}_t = \bs{W} \begin{pmatrix}
        1 \\ 
        \boldsymbol{\Psi}_{t,\cdot}
    \end{pmatrix},
\end{equation*}
where $\bs{W} \in \mathbb{R}^{n_h \times (p+1)}$ is a matrix containing the hidden weights. 

Crucially, the hidden weights are randomly drawn once and are fixed before the model training process, so the computational cost of estimating hidden layer weights via backpropagation is entirely avoided. Since a random-weight neural network can approximate any data-generating function with arbitrary accuracy \parencite{ELMFunctions}, it allows our model to learn nonlinear relationships between the response data's mixture model parameters and the environmental covariates. The element-wise logistic function is not a strict requirement; the same guarantees hold for any element-wise infinitely differentiable function. We chose the logistic function due to its ubiquity in both random and non-random neural network literature.

The approach taken in \textcite{flowmix} is a special case of this model when the original covariates are used as regressors instead of the principal components, $n_h = p$, $\boldsymbol{W} = \boldsymbol{I}_{p + 1}$, and the activation function $\sigma(\cdot)$ is set to be the identity function $\sigma(\boldsymbol{h}_t) = \boldsymbol{h}_t$. In Sections \ref{sec:sim} and \ref{sec:app} we compare our proposed model to one we call ``linear'', which is identical to the one developed by \textcite{flowmix} but uses principal components as regressors. This allows us to easily compare the interpretations of different models -- nonlinear and linear -- estimated using a common set of regressors. Also, using $q = 9$ principal components enables simpler interpretation than if we had used the original $p=37$ covariates. 

In all of our experiments, we set the hyperparameters $a = 0.5$ and $n_h = 70$. The $\mathrm{Unif}(-a, a)$ distribution used to draw the random weights in \eqref{eqn:hidden} only involves a single hyperparameter $a$, but can be replaced with any symmetric distribution centered at 0. We chose $a=0.5$ which induces roughly a $\mathrm{Unif}(0, 1)$ distribution on $h_{t, 1}, \ldots, h_{t, 70}$, and resulted in enough variation in the hidden layer weights to estimate a wide variety of regression functions. $n_h$ should be relatively large to allow flexibility in the model, although the marginal gain in accuracy diminishes at a certain point, and at the expense of increased computation. For choosing $n_h$ we recommend estimating models on a coarse $n_h$ grid (for example, $n_h = 30, 60, 90, \ldots$), plotting the negative log-likelihood against $n_h$, and choosing $n_h$ as the value at which the negative log-likelihood starts to plateau.

\subsection{Estimation} \label{sec:est}

Following \textcite{flowmix}, we estimate model parameters using the frequentist approach of maximizing a constrained pseudolikelihood. Phytoplankton flow cytometry data naturally have a class imbalance due to the nature of small, lightweight cells to dominate in number. Larger organisms are less numerous but still account for a significant portion of the biomass in their habitat \parencite{maranon2015}. Mixture models tend to overfit to the class which dominates in number and therefore underestimate the abundance of rare classes \parencite{Xu1996}, so we give additional importance to less numerous but more massive cells by exponentiating the likelihood contribution of each particle by its biomass. This results in the \textit{pseudo}likelihood:
\begin{equation}
\label{eqn:likelihood}
    \mathcal{L} (\alpha, \beta, \Sigma ; \{\boldsymbol y_{i}^{(t)}\}_{i, t}, \{ \boldsymbol{x}_t\}_t ) = \prod_{t = 1}^T \prod_{i = 1}^{n_t} \left[\sum_{k = 1}^K \pi_{k,t}(\boldsymbol{x}_t) \phi_d\left(\boldsymbol{y}_i^{(t)} | \boldsymbol{\mu}_{k, t} (\boldsymbol{x}_t), \boldsymbol{\Sigma}_k\right)\right]^{c_i^{(t)}}.
\end{equation}
Here, we use $\alpha$ to denote the collection of coefficients $\{\alpha_{0,k}\}_k, \{\boldsymbol \alpha_k\}_k$, and $\beta$ to denote $\{\boldsymbol \beta_{0,k}\}_k, \{ \boldsymbol \beta_k\}_k$. We also write $\pi_{k,t}(\boldsymbol{x}_t)$ and  $\boldsymbol{\mu}_{k, t} (\boldsymbol{x}_t)$ to emphasize their dependence on the covariates. Pseudolikelihoods are commonly utilized in informative sampling scenarios, where the likelihood contributions are exponentiated by the observations' inverse probabilities of selection. The resulting pseudolikelihood converges to the true population-generating distribution under certain regularity conditions \parencite{sav_toth_pseudo}. In our dataset, particles with large biomass tend to be less numerous and therefore less likely to be observed, so it is reasonable to assume that a particle's selection probability is approximately proportional to the inverse of its biomass. Figure S14 in Supplement G shows that the biomass weights are effective in improving the estimated distribution of biomass in the SeaFlow application. 

In addition, \textcite{flowmix} also impose two model constraints to encourage smoothness and penalize overly complex models, a technique we also utilize. First, since the optical properties of any particular species is limited in range due to biological constraints, we constrain the cluster means $\boldsymbol{\mu}_{k,t}$ to vary over time by up to a fixed amount $r$, called the maximum deviation. This constraint can be expressed as
\begin{equation} \label{eqn:maxdev}
    ||\boldsymbol{\mu}_{k, t} - \boldsymbol{\beta}_{0,k} ||_2 = ||\boldsymbol{\beta}_k' \sigma(\boldsymbol{h}_t) ||_2 \leq r,
\end{equation}
for cluster $k = 1, \ldots, K$ and time $t = 1, \ldots, T$. Another benefit of this constraint is that a properly chosen value of $r$ often prevents the switching of cluster labels over time. Since \eqref{eqn:maxdev} restricts the cluster means to a hypersphere, the data must be scaled so that each response dimension has roughly equal range. Second, to guard against selecting an overly complex regression model, lasso penalties \parencite{lasso} are imposed on all regression coefficients. Now, maximizing the constrained and penalized pseudolikelihood can also be formulated as the following optimization problem:
\begin{equation} \label{eqn:opt}
\begin{split} 
    \min_{\alpha, \beta, \Sigma} 
     - \frac{1}{N} \log \mathcal{L} (\alpha, \beta, \Sigma ; \{\boldsymbol y_{i}^{(t)}\}_{i,t}, \{ \boldsymbol{x_t}\}_t)
    + \lambda_\alpha \sum_{k = 1}^K ||\boldsymbol{\alpha}_k||_1 + \lambda_\beta \sum_{k = 1}^K ||\boldsymbol{\beta}_k||_1 \\ 
    \text{subject to } ||\boldsymbol{\beta}_k' \sigma(\boldsymbol{h}_t) ||_2 \leq r\ \forall k = 1, \ldots, K,\ \forall t = 1, \ldots, T,
\end{split}
\end{equation}
the solution of which provides parameter estimates. We solve \eqref{eqn:opt} using the penalized EM algorithm described in \textcite{flowmix} and use cross-validation to select the optimal hyperparameters $\lambda_\alpha$ and $\lambda_\beta$. To avoid data leakage between train and validation folds, we learn the PCA transformation from only the training data within each fold, though $q = 9$ is fixed \textit{a priori} as in Section \ref{sec:method}.

Lastly, since flow cytometry datasets have very large sample sizes $n_t$, we follow the approach taken by \textcite{flowmix} to reduce computational demand by aggregating particles into a fixed grid of $D$ bins in each cytogram dimension. The new binned cytogram at time $t$ consists of observations $\left\{\boldsymbol{y}_b^{(t)} : b = 1, \ldots, D^d\right\}$, where $\boldsymbol{y}_b^{(t)}$ is the center location of bin $b$.  We associate with bin $b$ a new weight $c_b^{(t)}$ equal to the total biomass of particles within the bin, and use $\boldsymbol{y}_b^{(t)}$ and $c_b^{(t)}$ in place of $\boldsymbol{y}_i^{(t)}$ and $c_i^{(t)}$ in \eqref{eqn:opt}. The total number of non-empty bins in the SeaFlow dataset is $N = \sum_{t = 1}^T n_t = 1,178,236$, which highlights the need for a computationally efficient nonlinear regression framework.

\section{Simulation} \label{sec:sim}

\subsection{Simulation design}
In this section, we generate pseudo-synthetic data to test the numerical properties of our model. For each signal size $\Delta$ and at each time point $t=1,\cdots, 296$, we generate $n_t = 1,000$ particles $\boldsymbol y_i^{(t)}$ from a mixture of two one-dimensional Gaussian distributions. Specifically, the latent membership $z_i^{(t)}$ is first drawn from $\mathrm{Categorical}(\pi_{1,t}, \pi_{2,t})$, and conditional on $z_i^{(t)}$, the particle's value $\boldsymbol y_i^{(t)} \mid z_i^{(t)}=k$ is drawn from $\mathcal{N}\left( \mu_{1,t}, 0.2^2\right)$ if $k=1$, or $\mathcal{N}\left( \mu^\Delta_{2,t}, 0.2^2\right)$ if $k=2$. Cluster 1's mean $\mu_{1,t}$ changes over time $t$ as a function of the two principal components $\bPsi_{\cdot, 1}$, and $\bPsi_{\cdot, 4}$, taken from the real data's principal component matrix $\boldsymbol{\Psi}$ in \ref{sec:data}. In contrast, cluster 2's mean $\mu_{2,t}^\Delta$ remains constant over time, so $\mu_{2,t}^\Delta = \mu_2^\Delta$ for all $t$. Also, cluster 1's cluster probability $\pi_{1, t}$ changes over time $t$ as a function of the two principal components $\bPsi_{\cdot, 1}$, and $\bPsi_{\cdot, 2}$, and $\pi_{2, t} = 1 - \pi_{1, t}$ by construction. 

There are three types of $\mu_{1,t}$ and $\pi_{1,t}$ configurations we consider:
\begin{enumerate}
\item {\bf Both linear.} Cluster means and logit-transformed cluster probabilities are both linear functions. \label{both-linear}
\item {\bf Nonlinear probability.} Fix the mean as a linear function and the logit-transformed cluster probabilities as nonlinear. \label{mean-nonlinear}
\item {\bf Nonlinear mean.}  
Fix the logit-transformed cluster probabilities as linear and means as nonlinear. \label{prob-nonlinear}
\end{enumerate}
For each configuration, we keep the first mean $\{\mu_{1,t}\}_{t=1}^{296}$ the same and vary the level of the second cluster mean $\mu_2^\Delta$ over signal sizes $\Delta$ which we define as the size of the gap between the two cluster means (averaged over time). That is, $\mu_2^0$ with the smallest signal size $\Delta=0$ has a zero gap between the average of the first cluster's mean $\frac{1}{T} \sum_{t = 1}^T \mu_{1,t}$ and $\mu_{2}^\Delta$. We consider equally spaced signal sizes between zero and the largest value $\Delta=0.95$, the signal size at which the two clusters' $\pm 2$ standard deviation bands around the mean cease to overlap.

We use three types of nonlinearity: an interaction, a quadratic function, and a standard logistic function. The exact functions we use to generate the data are as follows (for $t=1,\cdots, 296$):
\begin{itemize}
    \item Mean functions: 
    \begin{itemize}
        \item Linear: $\mu_{1,t} = 1 + 0.015 \bPsi_{t, 1} + 0.035 \bPsi_{t, 4}$
        \item Nonlinear:
        \begin{itemize}
            \item Interaction: $\mu_{1,t} = 1.05707 + 0.015 \bPsi_{t, 1} +  0.02 \bPsi_{t, 4} - 0.003 \bPsi_{t, 1} \bPsi_{t, 4}$ 
            \item Quadratic: $\mu_{1,t} = 1.030392 + 0.015 \bPsi_{t, 1} +  0.01 \bPsi_{t, 4} |\bPsi_{t, 4}| $
            \item Logistic:$\mu_{1,t} = 0.9396055+ 0.15 (1 + \exp(\bPsi_{t, 1}))^{-1} +  0.035 \bPsi_{t, 4}$
        \end{itemize}
    \end{itemize}
    \item Probability functions:
    \begin{itemize}
        \item Linear: $\text{logit}(\pi_{1,t}) = - 0.4 \bPsi_{t, 1} + 0.1 \bPsi_{t, 2}$
        \item Nonlinear:
        \begin{itemize}
            \item Interaction: $\text{logit}(\pi_{1,t}) = - 0.4 \bPsi_{t, 1}  + 0.02 \bPsi_{t, 1} \bPsi_{t, 2}$
            \item Quadratic: $\text{logit}(\pi_{1,t}) = - 0.4 \bPsi_{t, 1} - 0.05 {\bPsi_{t, 2}}^2$
            \item Logistic: $\text{logit}(\pi_{1,t}) = 2 -3.5 (1 + \exp(\bPsi_{t, 1}))^{-1} +  0.1 \bPsi_{t, 2}$.
        \end{itemize}
    \end{itemize}
\end{itemize}
These functions generate synthetic data which reproduce key characteristics of \textit{Prochlorococcus} and \textit{PicoEukaryote} observed in the Gradients 2 cruise, including realistic ranges of cell diameter and fluorescence intensities.

Next, for evaluating predictive out-of-sample accuracy, we use the log-pseudolikelihood in \eqref{eqn:likelihood}, calculated on an independently drawn dataset generated using the same parameters as the training dataset. For all scenarios, we use $5$-fold cross-validation to select tuning parameters, and for each model fit, we run the EM algorithm $30$ times. More specifically, we use a cross-validation scheme designed to prevent the data in the train and test folds from being too close to each other. Rather than using random splits of the time domain $\{1,\cdots, T\}$ for cross-validation folds, we  partition it into blocks of 20 consecutive time points and then distribute the blocks among the folds. Specifically, for $K = 5$ folds, the first fold is assigned blocks $1,6,11,\cdots$ and contain time points $\{1,\cdots, 20\}\cup\{ 121, \cdots, 140\} \cup \cdots$, the second fold is assigned the blocks $2,7,12, \cdots$, and so forth. 

\begin{figure}
    \centering
    \includegraphics[width=\linewidth]{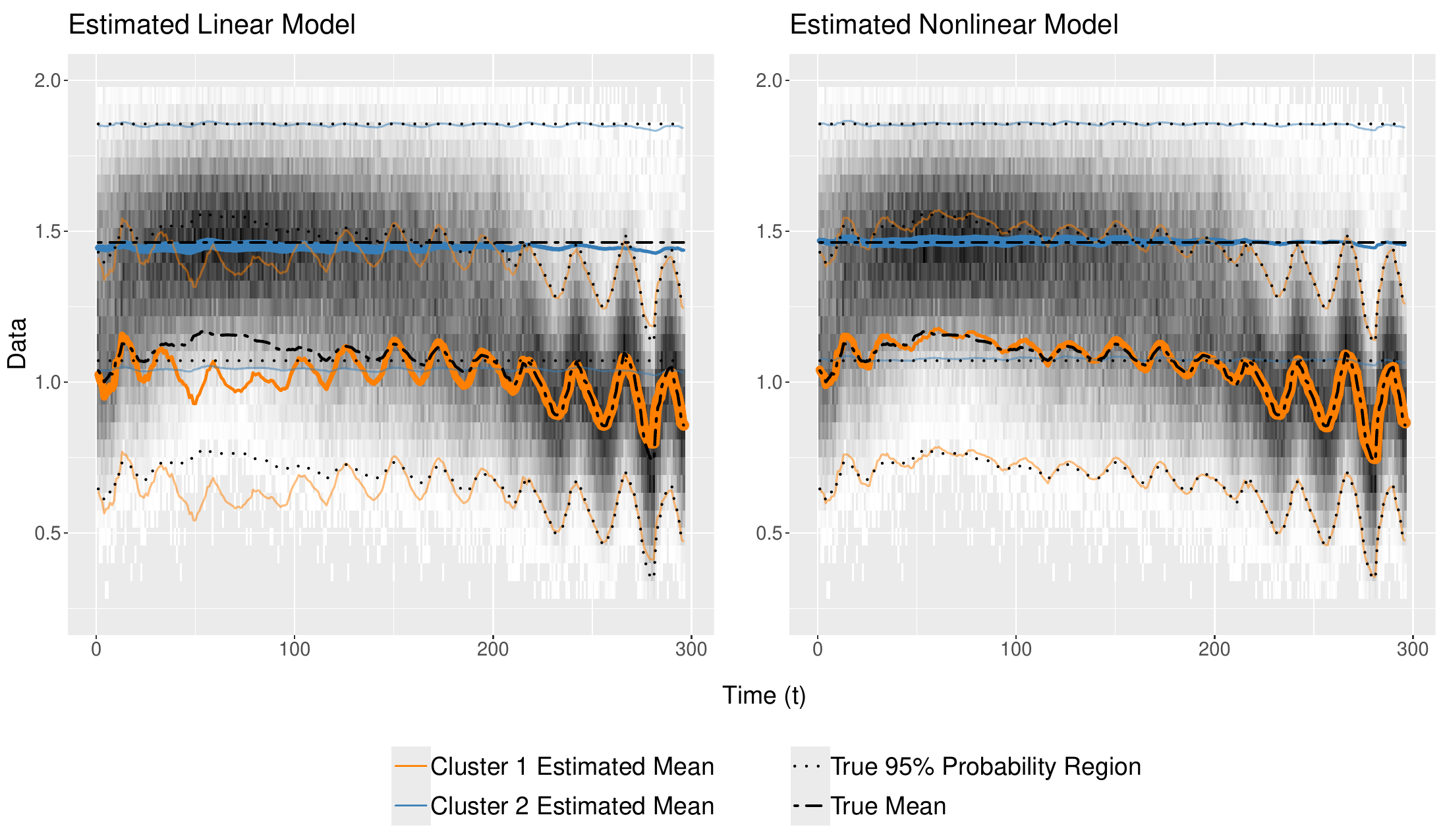}
    \caption{\textit{Estimated model on simulated 1-dimensional data. Both panels display in the background the simulated and binned data $y_b^{(t)}$ over time $t$, generated from the ``Interaction Mean'' model, with signal size $\Delta = 0.40384$. The greyscale color of the bin $b$ at time $t$ is proportional to the biomass $c_b^{(t)}$. Overlaid are the estimated parameters $\hat{\boldsymbol{\mu}}_{k, t}$ over time $t$ from the linear (left) and nonlinear (right) models as solid colored lines. These lines' thickness at time $t$ is proportional to the estimated cluster probability $\hat \pi_{k,t}$. The thin colored lines are the bounds of symmetric 95\% pointwise conditional probability regions of each Gaussian.
    The black dashed lines represent the true cluster means, and the black dotted lines represent the true 95\% central probability regions. 
    }}
    \label{fig:sim_data}
\end{figure}

\subsection{Simulation results}

Figure~\ref{fig:sim_data} shows an example of the estimated linear model (left panel) and nonlinear model (right panel), both using data generated from a linear probability function and an interaction mean function. In the true model, the variation in the first cluster mean $\{\mu_{1, t}\}_{t=1}^{296}$ over time is largely due to $\bPsi_{\cdot, 4}$, which is highly correlated with sunlight and thus oscillates regularly with a daily periodicity. The interaction in the linear effect of $\bPsi_{\cdot,4}$ with $\bPsi_{\cdot,1}$ causes the true amplitude of the oscillation to become smaller at high latitudes. However, the estimated linear model fails to capture this and the amplitude of the estimated oscillation in ${\mu}_{1, t}$ does not shrink enough in high latitudes. It also noticeably underestimates the first cluster mean $\mu_{1,t}$ when $t \in [40, 110]$, as well as the second cluster mean $\mu_{2,t}$ across all time points. In contrast, the nonlinear model more accurately captures the change in amplitude, with its estimated means closely matching the ground truth at all time points. Supplement E shows that the nonlinear model results are reproducible across a variety of hidden layer widths (Figure S11) and draws of the random weights (Figure S12).

\begin{figure}
    \centering
    \includegraphics[width=\linewidth]{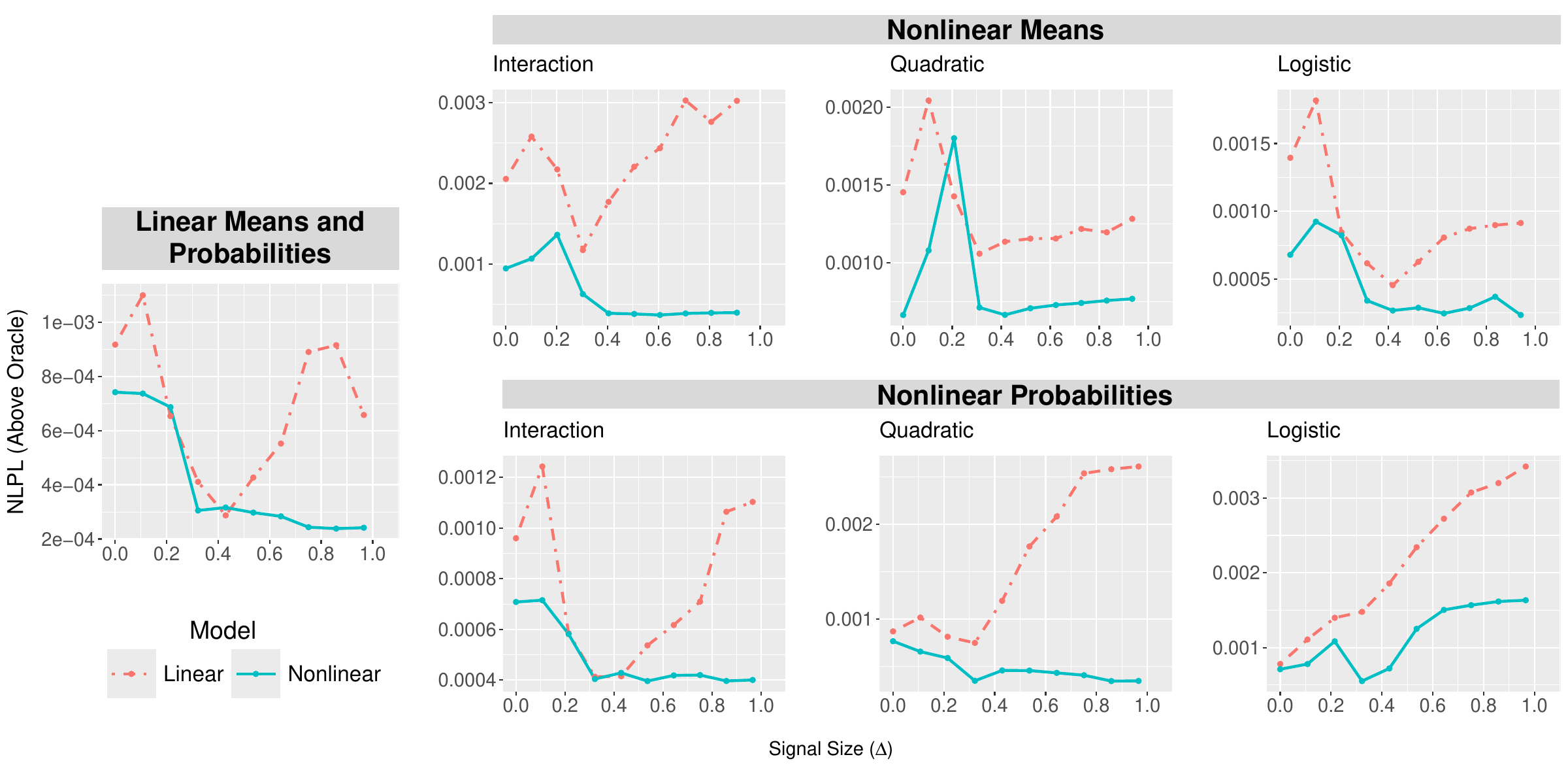}
    \caption{\textit{Out-of-sample negative log-pseudolikelihoods (NLPLs) after subtracting the NLPL of the oracle of each model at each signal size. The linear model is represented by the dashed line and the nonlinear by the solid line.}}
    \label{fig:nll}
\end{figure}

Figure~\ref{fig:nll} shows the aggregate summary of the simulation results. The y-axis values in each panel are the negative log-pseudolikelihoods (NLPLs) of each estimated model {\it after} subtracting the NLPL of the oracle, or true generating model. It is clear that the nonlinear model almost uniformly outperforms the linear model in terms of the out-of-sample negative log-pseudolikelihood, which demonstrates a better fit to the data. (The nonlinear model even outperforms for linear data sometimes, which may be due to large sample sizes with relatively little noise, which makes training and test datasets very similar.) We also provide a simulation with three-dimensional response data in Supplement D, which highlights the ability of our proposed model to recover multivariate Gaussian mixtures under heavy cluster overlap. The model fit to the data is shown in Figure S8, and a comparison of the true and estimated parameters is provided in Figure S9. In addition, we compare the performance of our model to that of alternative methods, including generalized additive models \parencite{Wood2017}, random forests \parencite{Breiman2001}, and mixture of experts distributional regression \parencite{Rugamer2024}. Figure S10 compares the parameter estimation accuracy among the considered models.

\section{Application to SeaFlow cruise data} \label{sec:app}

In this section, we estimate both the linear model and the nonlinear model on the real flow cytometry dataset described in Section \ref{sec:data}. We bin the cytograms $\boldsymbol y^{(t)}$ with a grid size of $D = 40$ bins in each dimension and set bin weights $c_b^{(t)}$ equal to each bin's total biomass. We followed the same cross-validation scheme as described in Section \ref{sec:sim} with $10$ EM algorithm restarts. 

\begin{figure}
    \centering
    \includegraphics[width=\linewidth]{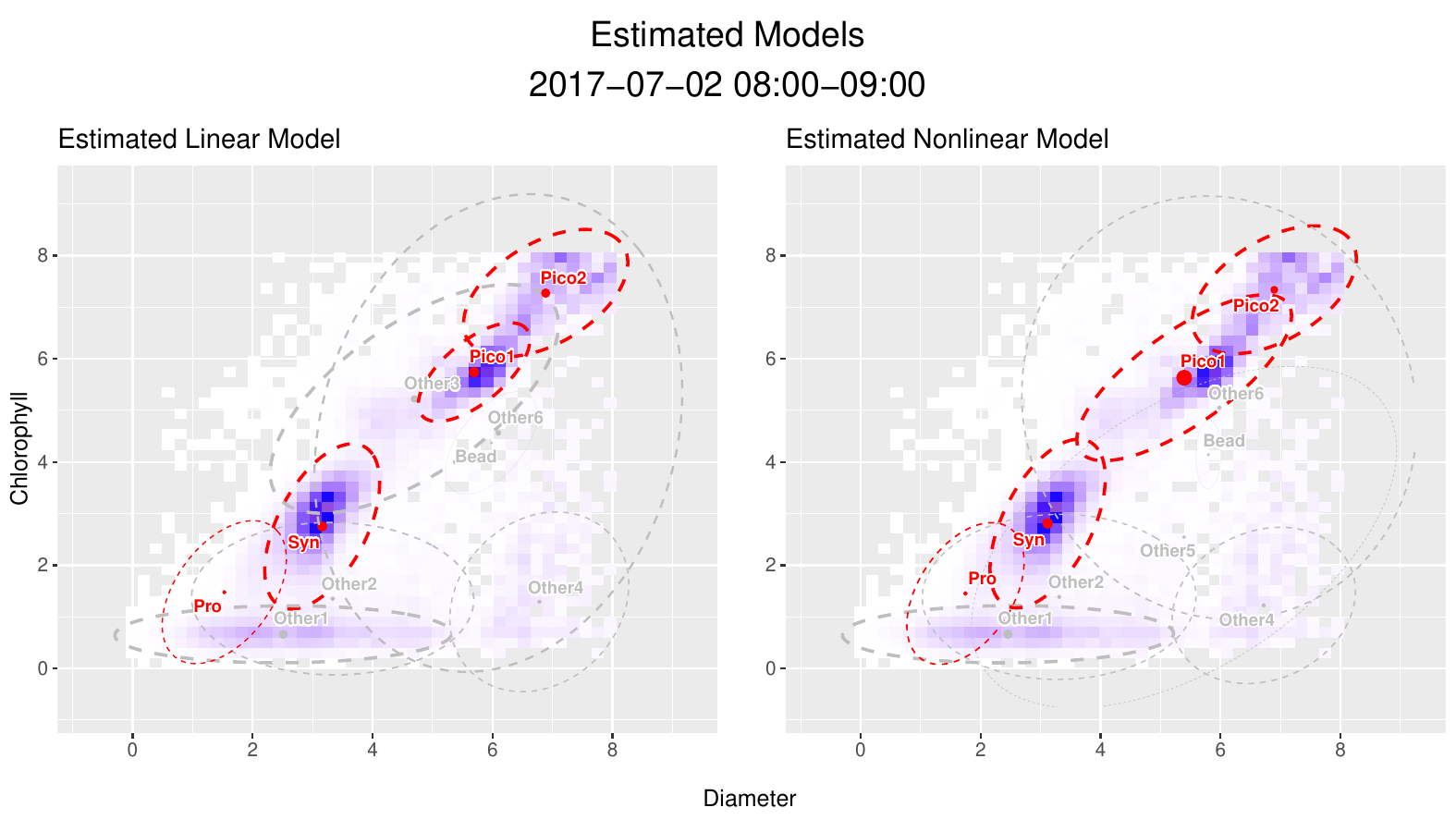}
    \caption{\textit{Cytogram in two dimensions, diameter and chlorophyll, at time $t = 33$ (2017-07-02 08:00-09:00), with model estimates overlaid in red (marking the important clusters) or grey (the rest). The data in the background are binned, with the intensity of blue hue proportional to the total biomass in each bin. The solid points mark the cluster means, and the sizes of the points are proportional to the estimated cluster probabilities. The ellipses represent 2-dimensional projections of symmetric cluster-conditional 95\% probability sets of each cluster.}}
    \label{fig:clusters33}
\end{figure}

The models' estimates at the time point $t = 33$ are shown in Figure~\ref{fig:clusters33},
where we focus on two out of the three cytogram dimensions, and overlay the estimated model with the binned data. The resulting clusters from each model have subsequently been expert-identified as \textit{Prochlorococcus} (Pro), \textit{Synechococcus} (Syn), and two distinct \textit{PicoEukaryote} (Pico1 and Pico2), as well as other populations. It is visually clear that the two models identify the major populations (Pro, Syn, Pico1 and Pico2 in Figure~\ref{fig:clusters33})  similarly. The nonlinear model predicts the \textit{PicoEukaryote 1} cluster to be bigger, which captures more of the biomass between the \textit{Synechococcus} and \textit{PicoEukaryote 2} clusters.

We also measured the out-of-sample accuracy of each model, using the following nested cross-validation scheme:
\begin{enumerate}
    \item Partition the time points $\mathcal{T} = \{1, \ldots, T\}$ into folds $\mathcal{T}^{(j)}$, $j = 1, \ldots, 5$, using equally spaced blocks of 20 consecutive time points. Partition the data accordingly as $\mathcal{D}^{(j)} = \left\{\left(\boldsymbol{y}_i^{(t)}, \boldsymbol{x}_t\right) : t \in \mathcal{T}^{(j)}\right\}$, for fold $j = 1, \ldots, 5$. 
    \item For outer fold $j = 1, \ldots, 5$:
    \begin{enumerate}
        \item Use $\mathcal{D}^{(j)}$ as a held-out dataset. Using the remaining data $\bigcup_{k \neq j} \mathcal{D}^{(k)}$, cross-validate the model to select the optimal hyperparameters, then refit the model as described in Section \ref{sec:est}, resulting in model parameter estimates $\hat \alpha$, $\hat \beta$, and $\hat \Sigma$. 
        \item Evaluate the unpenalized negative log-pseudolikelihood $-\mathcal{L} \left(\hat{\alpha}, \hat{\beta}, \hat{\Sigma} ; 
        \{\boldsymbol{y}_i^{(t)}\}_{i, t \in \mathcal{T}^{(j)}}, \{\boldsymbol{x}_t\}_{t \in \mathcal T^{(j)}}\right)$.
    \end{enumerate}
    \item Let $-\frac{1}{5} \sum_{j = 1}^5 \mathcal{L} \left(\hat{\alpha}, \hat{\beta}, \hat{\Sigma} ; 
        \{\boldsymbol{y}_i^{(t)}\}_{i, t \in \mathcal{T}^{(j)}}, \{\boldsymbol{x}_t\}_{t \in \mathcal T^{(j)}}\right)$ be the final estimate of the out-of-sample model fit.
\end{enumerate}

\begin{figure}
    \centering
    \includegraphics[width=\linewidth]{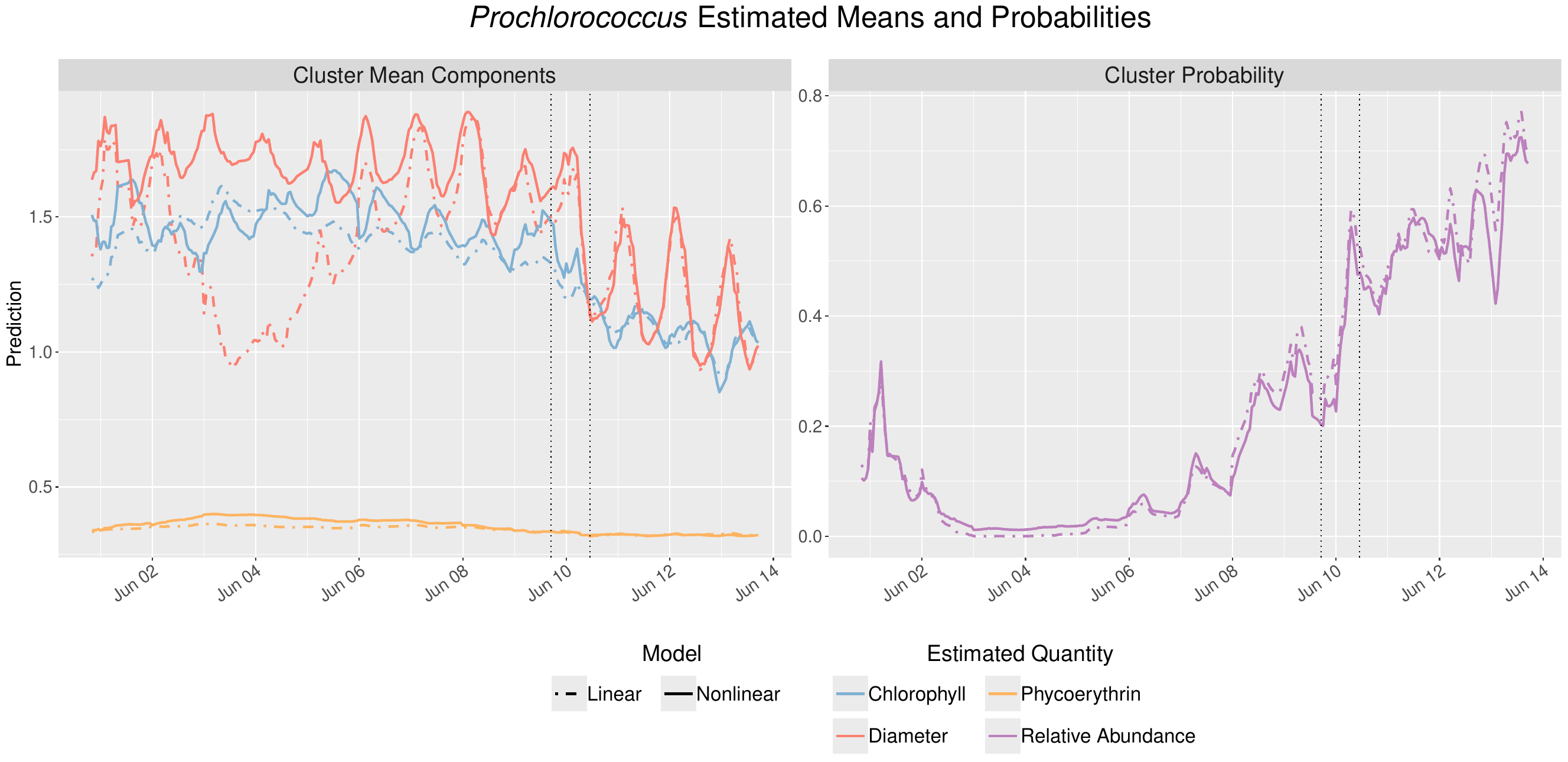}
    \caption{\textit{Mean and probability estimates over time for the \textup{Prochlorococcus} cluster. In the left panel, the three dimensions of the cluster mean are plotted together; diameter, chlorophyll, and phycoerythrin estimates are represented by the red, blue, and yellow lines, respectively. In the right panel, relative abundance predictions are displayed. The nonlinear model predictions are represented by the solid lines and the linear model predictions by dot-dashed lines. The vertical dotted lines represent the boundaries of the North Pacific Transition Zone. Figures S3–-S5 in Supplement B present similar plots for \textup{Synechococcus} and \textup{PicoEukaryote} clusters 1 and 2, respectively.}}
    \label{fig:gating}
\end{figure}

The linear model had an out-of-sample negative log-pseudolikelihood of $3.486$ while the nonlinear model's was $3.414$, showing similar generalization performance. In addition, the relative abundance predictions of the two models are very similar, as Figure~\ref{fig:gating} shows. However, there are notable differences between the two models' cluster mean predictions. The linear model predicts a dramatic decrease in \textit{Prochlorococcus} diameter between July 2nd and July 6th, while the nonlinear model predictions remain relatively stable, showing a diel oscillation at a similar overall level. Also, the nonlinear model better captures the amplitude of the expected diel oscillation in chlorophyll of {\it Prochlorococcus} throughout all time points, matching its expected behavior from the corresponding oscillation in cell size. Overall, while the models display similar clustering results, the nonlinear model estimates more flexible cluster mean movement over time that fits scientific intuition.

Next, we interpret the estimated models in relation to the principal components of the environmental covariates. Since we use a neural network with random weights, the parametric form of the estimated regression equations is difficult to interpret analytically. Instead, we interpret the relationship between the parameters $(\boldsymbol{\mu}_{k, t}, \pi_{k, t})$ and principal components $\boldsymbol{\Psi}_{t, j}$ visually using partial response curves. We define partial response curves as the average of a model parameter estimate given a particular value of principal component $j$ and average values of all other principal components, constructed as follows. For example, let us denote as:
\begin{equation*}
\hat{\boldsymbol{\mu}}_{k}(\psi_1, \cdots, \psi_q) = \boldsymbol{\hat \beta}_{0,k} + \boldsymbol{\hat \beta}_k' \sigma(\boldsymbol W\begin{psmallmatrix}1\\ \psi_1\\ \vdots\\\psi_q\end{psmallmatrix}), 
\end{equation*}
the $k$th cluster mean predictions evaluated at scalar principal component values 
$\psi_1, \cdots, \psi_q$. Also denote $\bar \bPsi_{\cdot ,l}$ as the time-wise average of the $l$th principal component. Then, the partial response curve of the $k$th cluster mean $\boldsymbol{\mu}_{k}$ in response to changes in the $j$th principal component is represented as 2-dimensional points:
$$(x,\; \hat{\boldsymbol{\mu}}_{k}(\bar\bPsi_{\cdot, 1}, \cdots, \bar \bPsi_{\cdot, j-1}, x, \bar \bPsi_{\cdot, j+1}, \cdots, \bar \bPsi_{\cdot, q})),$$
calculated over a fine grid of values of $x$ in the range of $j$th principal component values $\bPsi_{\cdot, j}$ in the data. Partial response curves for relative abundance $\pi_k$ are defined similarly.

\begin{figure}
    \centering
    \includegraphics[width=\linewidth]{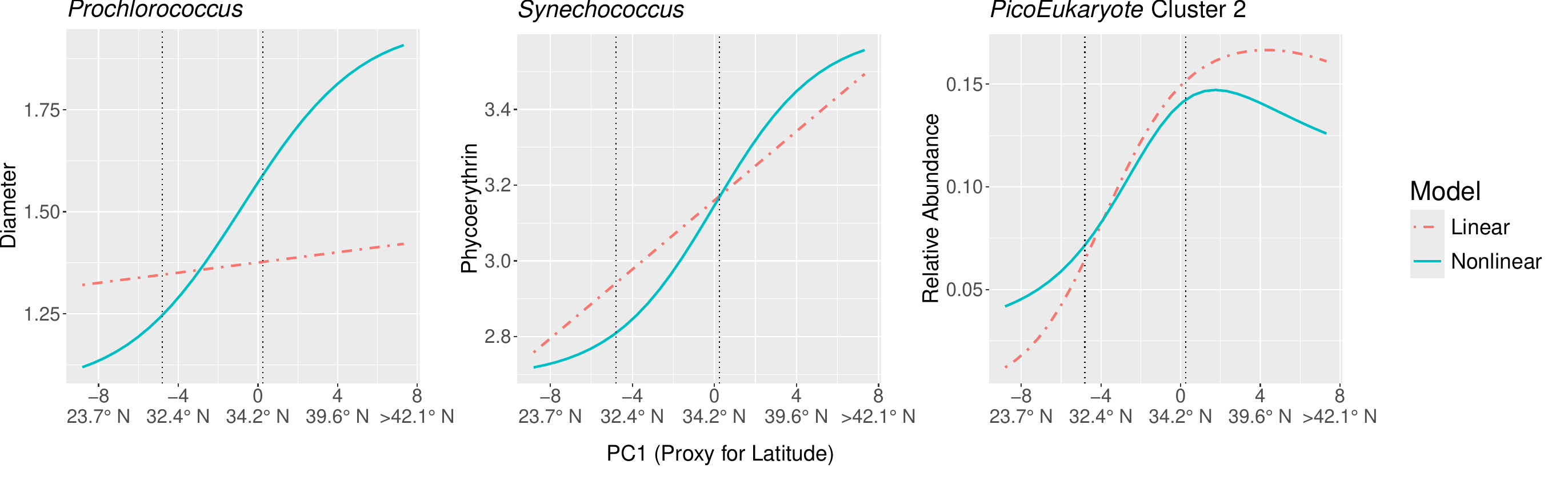}
    \caption{\textit{Partial responses of \textup{Prochlorococcus} diameter, \textup{Synechococcus} phycoerythrin fluorescence, and the second \textup{PicoEukaryote} cluster's relative abundance as functions of the first principal component. The vertical dotted lines represent the boundaries of the North Pacific Transition Zone. The full set of partial response curves for PC1 is provided in Figure S6 (Supplement C).}
    }
    \label{fig:PC1_PDP}
\end{figure}

Next, as a visual aid for interpreting the results, we plot two vertical dotted lines in each panel of Figure~\ref{fig:gating} and Figure~\ref{fig:PC1_PDP} representing the approximate start and end of the North Pacific Transition Zone (NPTZ), estimated directly from the data (bottom-left panel of Figure~\ref{fig:pc_cov_corr}). These lines mark a region where ocean conditions rapidly change latitudinally.

We see in Figure~\ref{fig:PC1_PDP} a few representative partial response plots in which certain phytoplankton properties are displayed as functions of PC1. As PC1 increases, the cruise moves northward closer to the subarctic region. We see in the left panel of Figure~\ref{fig:PC1_PDP} that both models predict \textit{Prochlorococcus} diameter to be larger monotonically as one moves to higher latitudes, if one were to hold as constant ocean water conditions that are not associated with latitude (PC1). This strongly suggests that the average \textit{Prochlorococcus} microbe is larger in higher latitudes. However, the linear model predicts only a flatter and linear change, by design. The nonlinear model is capable of predicting a more flexible change. It predicts \textit{Prochlorococcus} diameter to change (1) much slower in the south-most and north-most regions along the cruise track, and (2) much faster in the transition zone, especially within the NPTZ. This is consistent with external studies \parencite{Thompson2021} that different \textit{subtypes} of \textit{Prochlorococcus} are dominant in each of the two gyres -- Subpolar and Subtropical -- and our model estimates further suggest that those subtypes have a clear size difference. 

For \textit{Synechococcus}, both models predict that phycoerythrin fluorescence increases monotonically as one goes from south to north along the cruise trajectory. As with \textit{Prochlorococcus} diameter, the nonlinear model curve flattens at extreme latitudes, indicating stable biological function of \textit{Synechococcus} cells. The linear model predictions, however, do not level off, and any prediction outside of the observed latitude range is sure to be too extreme to be scientifically useful. This finding is especially important since the presence of phycoerythrin in a cell uniquely identifies it as \textit{Synechococcus}, since other phytoplankton contain little to none of this pigment.

Finally, the second \textit{PicoEukaryote} cluster (Pico2) is predicted to have a maximum relative abundance at a value of PC1 which corresponds to latitude 38.8$^\circ$ N. At more northern locations than this, the relative abundance is predicted to decrease. Such results highlight the possibility that nonlinear mixture-of-experts models, such as our proposed model,  can be used to learn globally and locally \textit{optimal} ocean conditions where certain phytoplankton species are most relatively abundant. In Supplement F, we refit the nonlinear model for different numbers of principal components; Figure S13 shows that the partial response curves remain relatively stable. We conclude that model interpretation is robust to the number of principal components input to the neural network.

\begin{figure}
    \centering
    \includegraphics[width=\linewidth]{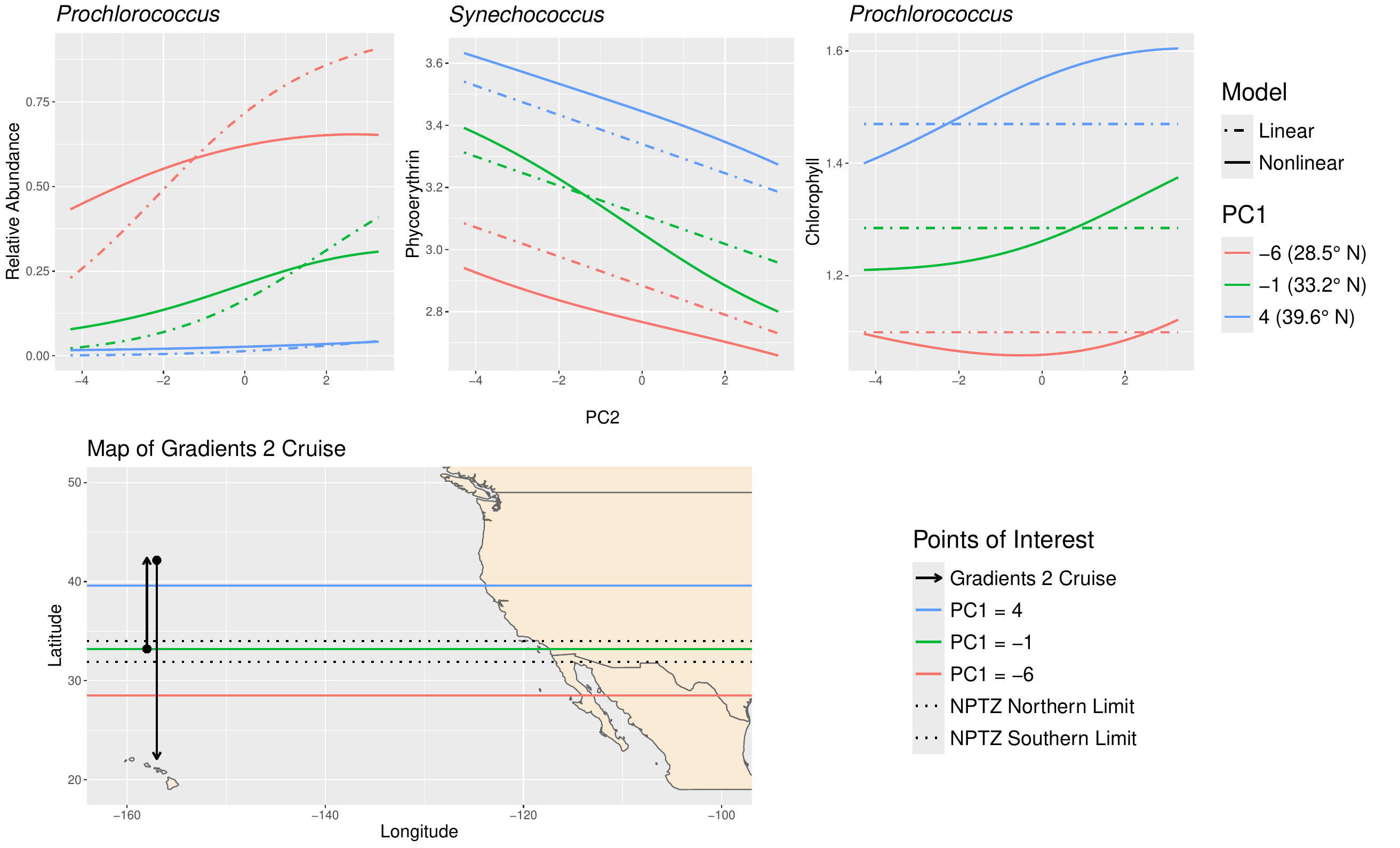}
    \caption{\textit{The top panel displays partial response curves of certain model parameters as functions of the second principal component, conditional upon several different values of the first principal component. The bottom panel displays the trajectory of the Gradients 2 cruise, with horizontal lines marking the values of PC1 used in the top panel as well as the NPTZ boundaries. The full set of partial response curves for PC2, conditional on PC1, is provided in Figure S7 (Supplement C).}}
    \label{fig:PC2_PDP}
\end{figure}

The nonlinear model also captures interaction effects between PC1 and PC2, as displayed in Figure~\ref{fig:PC2_PDP}.  In each panel, the three lines show a {\it conditional} partial response curve of the nonlinear model's parameter, say $\theta$, defined by first fixing PC1 to be a certain value $c$, then plotting the partial response of the parameter estimate $\hat \theta$:
$$(x, \;\hat \theta(c, x, \bar \bPsi_{\cdot, 3}, \cdots, \bar \bPsi_{\cdot, q}))$$
over a fine grid of values of $x$ in the range of the second principal component $\bPsi_{\cdot, 2}$. 
Figure~\ref{fig:PC2_PDP} highlights the conditional partial responses of the cluster probability $\hat \pi_{k,t}$ (in the left panel) or one dimension of the cluster mean $\hat{\boldsymbol{\mu}}_{k,t}$ (in the center and right panel), fixing the values of PC1 at three values ($-6$, $-1$, and $4$) which correspond to three locations clearly south of, inside of, and north of the North Pacific Transition Zone.

First, for cluster means, whenever the three lines from the same model are not parallel and equidistant, this indicates an interaction between PC1 and PC2. We see this most clearly in \textit{Prochlorococcus} chlorophyll fluorescence in the right panel. At a low value of PC1 (-6, corresponding to south of the NPTZ), the partial response curve is relatively flat, convex, and quadratic. At PC1 = -1 (inside the NPTZ), fluorescence increases monotonically at a faster rate, and at PC1 = 4 (north of the NPTZ) the curve becomes concave, meaning fluorescence increases at a decreasing rate. Since the linear model cannot estimate variable rates of change, it averages the increasing and decreasing effects, so fluorescence is estimated to be constant in all three regions. In the center panel showing the phycoerythrin fluorescence of \textit{Synechococcus}, the partial response curve shows a much faster rate of decrease within the transition zone than outside it.

Next, we examine the relative abundance prediction of \textit{Prochlorococcus}
(left panel of Figure~\ref{fig:PC2_PDP}). First, the linear model's partial response curve to PC2 follows a similar logistic shape at all specified values of PC1. On the other hand, the nonlinear model curve clearly changes in shape, especially from PC1 = $-1$ to PC1 = $-6$.  This suggests an interaction effect between PC1 and PC2 which allows the response of \textit{Prochlorococcus} relative abundance to PC2 to be much more flexible depending on the latitude, of which PC1 serves as a proxy.  Furthermore, the nonlinear model has a shallower slope overall, thus estimating \textit{Prochlorococcus} abundance
to be more stable in response to physical oceanographic attributes (PC2) in southern waters than the linear model does.

\section{Discussion} \label{sec:disc}

In this article we propose a novel mixture of neural network experts which uses randomly generated and fixed hidden layer weights to represent a nonlinear response to the covariates. This allows one to capture nonlinear behavior in flow cytometry datasets with the key advantage of a relatively low computational cost. We also use principal components as regressors to allow for compression of environmental covariates. When paired with partial response plots, using principal component regressors enhances our ability to interpret the model with respect to major modes of variation in the environmental factors. We conduct an extensive simulation to show that the proposed model matches the clustering performance of similar linear models and more accurately fits nonlinear data. Also, we apply our model to real phytoplankton flow cytometry data collected along a north-south transect on the North Pacific Ocean. The results show that the nonlinear model better reflects expected behaviors of phytoplankton traits when compared to the linear model. For example, the nonlinear model predicts the rate of change in cells' optical properties and abundances to depend on how rapidly the environmental conditions change. Furthermore, optical properties are predicted to level off when environmental conditions are at very low or high values, consistent with the fact that microbes of a certain species have a biological upper and lower limit to their size. This saturation behavior represents a fundamental biological constraint that linear models cannot capture, as they assume constant rates of change across the entire environmental gradient. The nonlinear approach allows different expert components to become active at different portions of the gradient, enabling the model to capture rapid responses under moderate conditions while transitioning to asymptotic behavior at environmental extremes. We therefore recommend the use of a nonlinear model when analyzing flow cytometry samples collected over a range of locations with diverse environmental conditions.

There is room for further model development. Our method assumes clusters are Gaussian, yet cell clusters in cytograms often exhibit skewness and heavy tails. For this reason, it is worthwhile to explore modifications with skew Normal or skew $t$ kernels as well as nonparametric kernel specifications, such as infinite mixtures. In fact, a Bayesian semiparametric extension may provide a better fit to the data with the added benefits of uncertainty quantification and hyperparameter tuning without resampling procedures or cross-validation. Another area for future work relates to the hardware limitations of flow cytometers; the Gradients 2 dataset contains some observations which are censored at SeaFlow's measurement limits. Proper handling of censored data may strengthen model validity and accuracy. Additionally, in the oceanographic setting, the responses are trivariate, whereas, in biomedical settings, the dimension is often much larger. This may introduce numerical instabilities and computational bottlenecks in estimating the covariance matrix of each Gaussian mixture component. In these cases, sparse precision matrix estimation techniques \parencite{Meinshausen2006,Friedman2007} may be necessary. We also note that PCA does not induce sparsity in the original variables. The use of a method such as sparse PCA \parencite{Jolliffe2003,Zou2006SPCA} or group lasso \parencite{Yuan2005} may provide sparser and therefore more easily interpretable latent factors. Our experiment in Supplement F also hints that some latent factors may contribute to little variability in the covariate space, but predict the response data well. This motivates an extension of our model which learns supervised latent factors, similar in spirit to partial least squares regression \parencite{Wold2001}. The correlated design and the potential for bias in the lasso estimator also warrants consideration of alternative coefficient penalties, including the trace lasso \parencite{Grave2011}, adaptive lasso \parencite{Zou2006AdaLasso}, and smoothly clipped absolute deviation penalty \parencite{Fan2001}. Finally, it would be valuable to extend our application to a larger collection of cruise datasets exploring an even wider range of locations and times. Mapping the spatial distributions and average optical properties of phytoplankton regionally or globally based on \textit{in situ} data, uncovering temporal trends, and identifying the environmental conditions which are most predictive of phytoplankton behavior, are ambitious yet exciting possibilities.

\section*{Acknowledgements}

The authors acknowledge the Hummingbird Computational Cluster at the University of California, Santa Cruz, for providing computing resources that have contributed to the research results reported within this publication. URL: \url{https://hummingbird.ucsc.edu/}.

\section*{Data Availability Statement}

The data that support the findings of this study are openly available in MoE-RWNN at \url{https://github.com/ethan-pawl/MoE-RWNN}.

\section*{Funding Statement}

This work was supported through grants by the Simons Collaboration on Computational Biogeochemical Modeling of Marine Ecosystems/CBIOMES (Grant ID: 1195553 to Sangwon Hyun), the Simons Microbial Oceanography Project Award (Grant ID: 574495 to Francois Ribalet), and the National Science Foundation awards NCSE-2215169, RISE-2530447, and RISE-2530448. 

\section*{Conflict of Interest Statement}

The authors declare no conflicts of interest.



\newpage

\printbibliography
\end{refsection}

\newpage 

\begin{refsection}

\begin{center}
{\LARGE{\bf Supplement to ``Mixtures of Neural Network Experts with Application to Phytoplankton Flow Cytometry Data''}}

\end{center}

\baselineskip=12pt

\vskip 2mm

\begin{center}

\setcounter{footnote}{0}

Ethan Pawl\footnote{\baselineskip=10pt Department of Statistics, University of California, Santa Cruz, 1156 High St.,
  Santa Cruz, CA 95064, epawl@ucsc.edu}
Fran\c{c}ois Ribalet\footnote{\baselineskip=10pt School of Oceanography, University of Washington, Seattle, 1410 NE Campus Pkwy.,
  Seattle, WA 98195,
  ribalet@uw.edu}
Paul A. Parker\footnote{\baselineskip=10pt Department of Statistics, University of California, Santa Cruz, 1156 High St.,
  Santa Cruz, CA 95064, paulparker@ucsc.edu}\, and
Sangwon Hyun\footnote{\baselineskip=10pt Department of Statistics, University of California, Santa Cruz, 1156 High St.,
  Santa Cruz, CA 95064,
  sangwonh@ucsc.edu}

\end{center}

\section*{Supplement A: Principal Components over Time}

\setcounter{figure}{0}
\setcounter{page}{1}
\renewcommand{\thefigure}{S\arabic{figure}}

\begin{figure}[H]
    \centering
    \includegraphics[width=.65\linewidth]{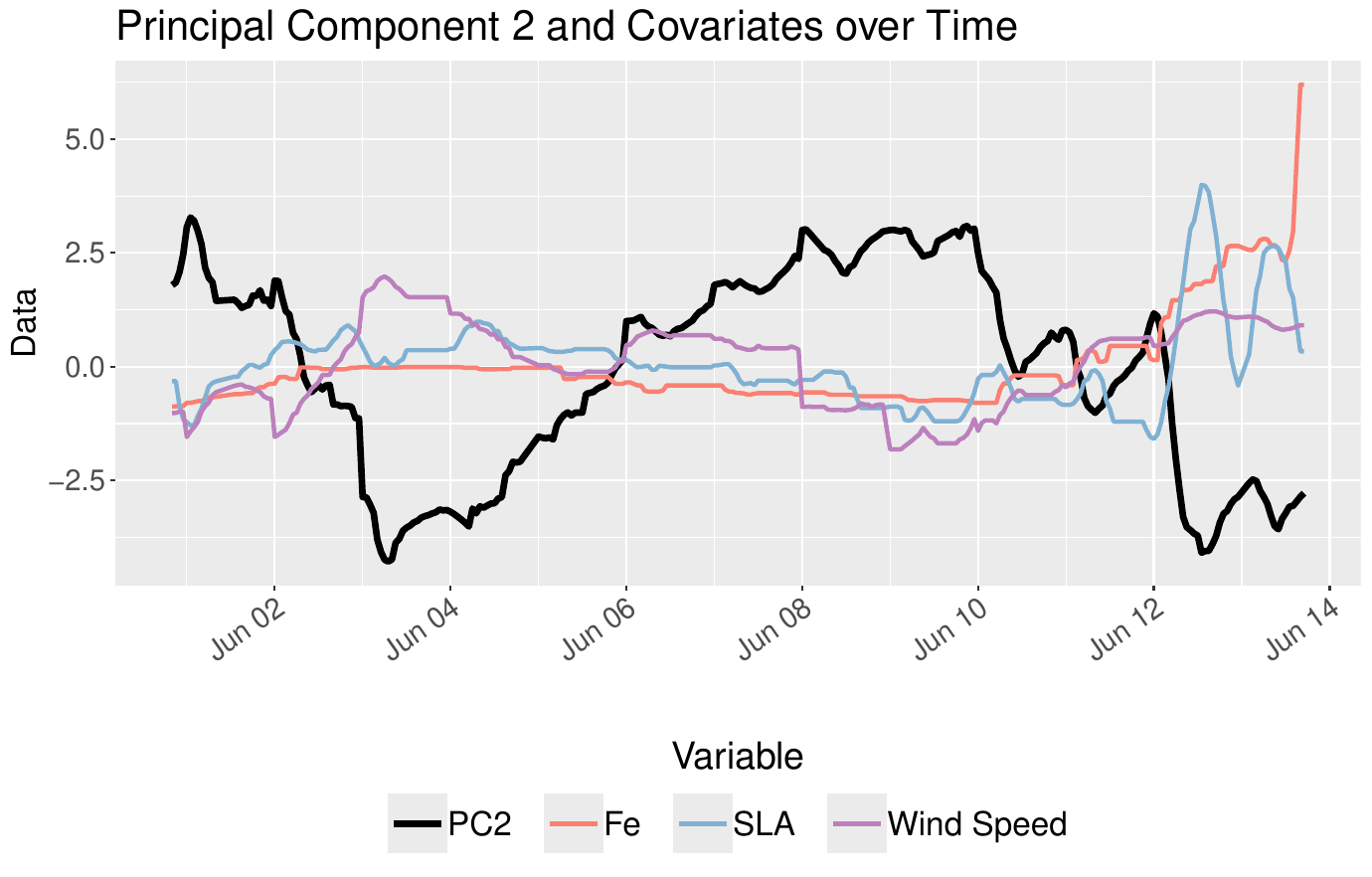}
    \caption{\textit{Principal component 2 and its most important constituent covariates over time: iron (Fe), sea level anomaly (SLA) and wind speed.}
    }
    \label{fig:PC2_line}
\end{figure}

\vskip 4mm
\baselineskip=12pt 

\baselineskip=12pt
\par\vfill\noindent
\par\medskip\noindent

\clearpage\pagebreak\newpage 
\baselineskip=24pt

\begin{figure}[H]
    \centering
    \includegraphics[width=\linewidth]{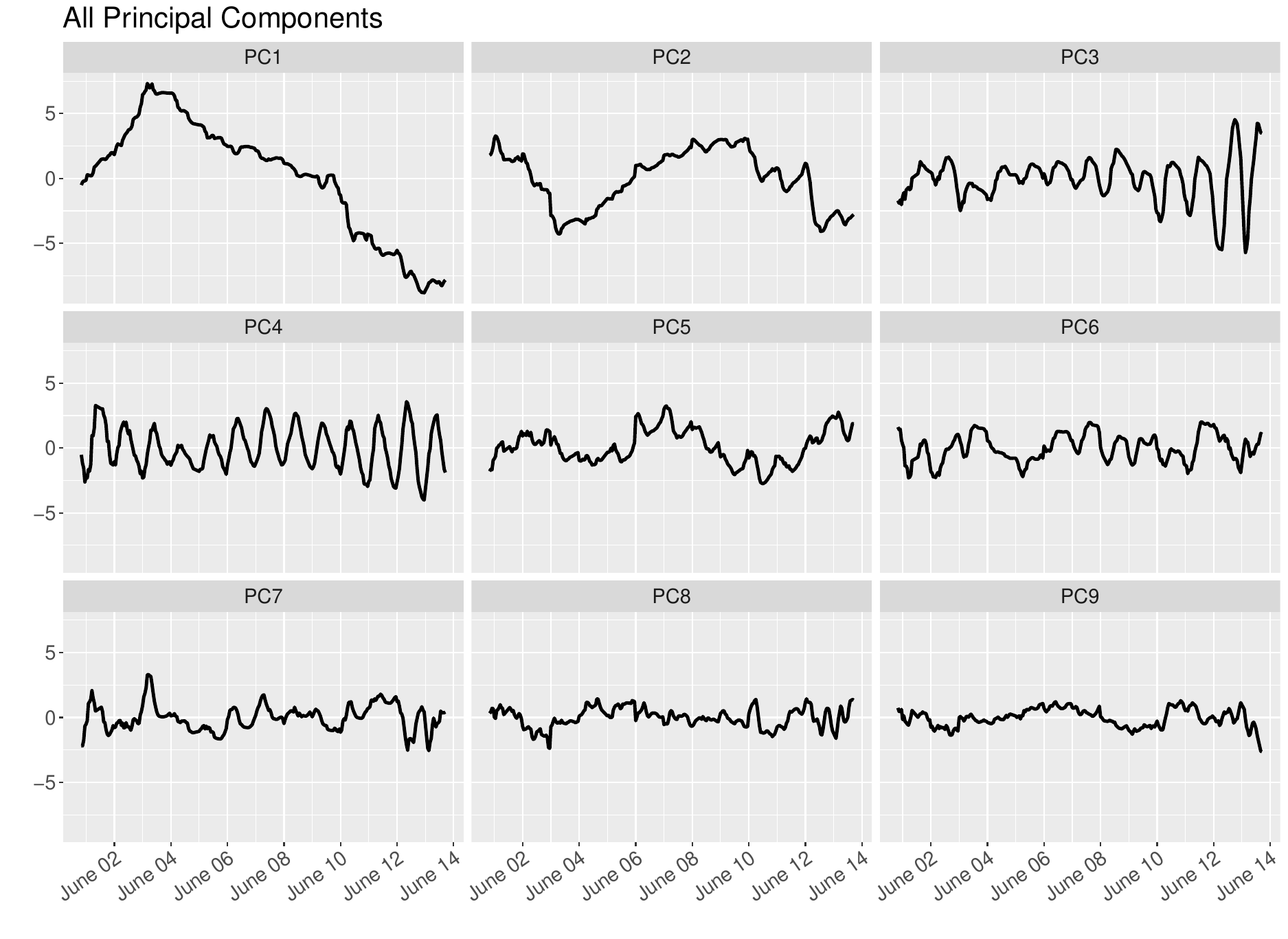}
    \caption{\textit{All nine principal components used in our data analysis.}}
    \label{fig:all_pcs}
\end{figure}

\section*{Supplement B: SeaFlow Cruise Data Clustering Results}

\begin{figure}[H]
    \centering
    \includegraphics[width=\linewidth]{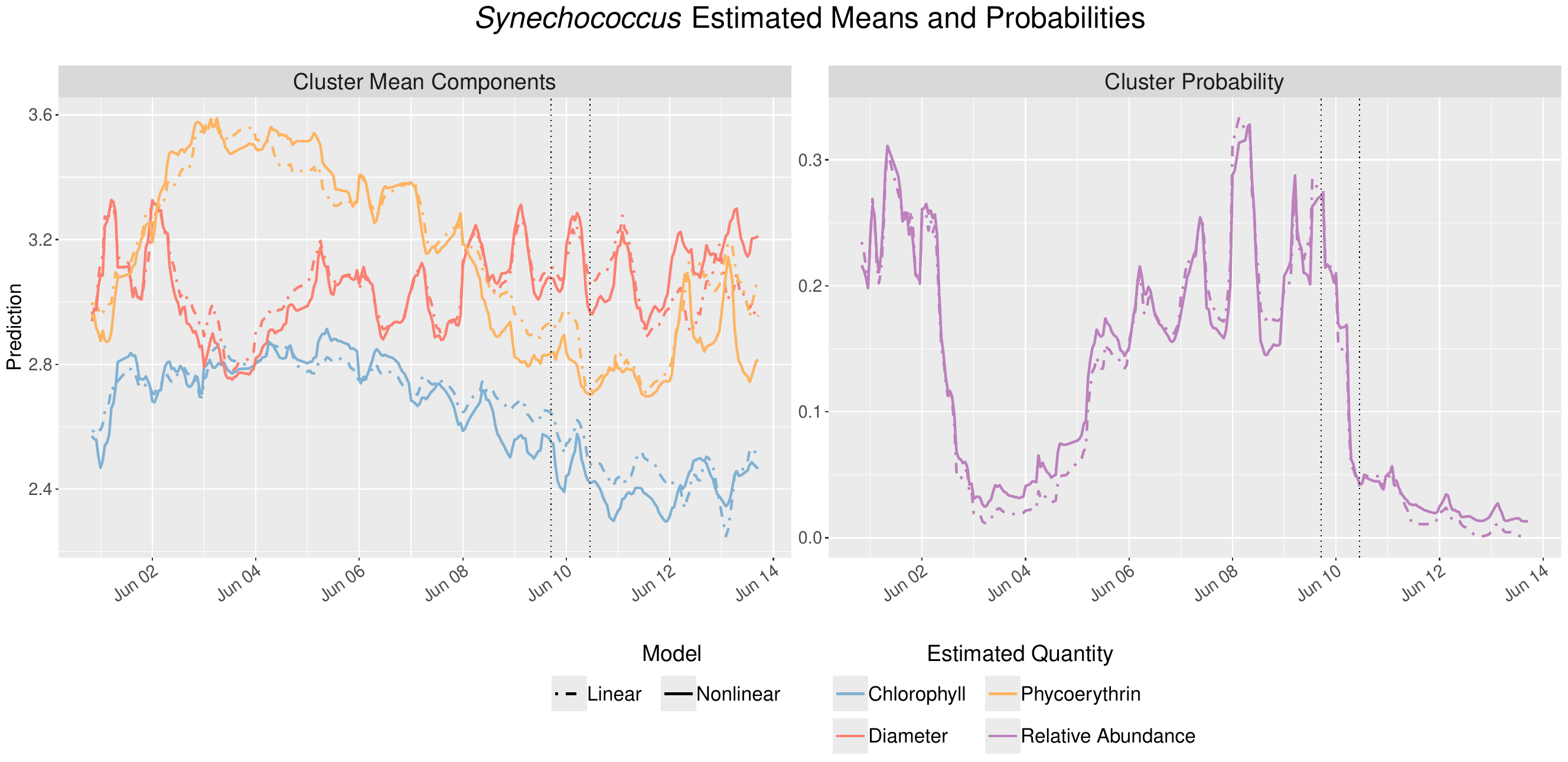}
    \caption{\textit{Estimated means and probabilities over time of \textup{Synechococcus} for each model.}}
    \label{fig:Sim_gate_syn}
\end{figure}

\begin{figure}[H]
    \centering
    \includegraphics[width=\linewidth]{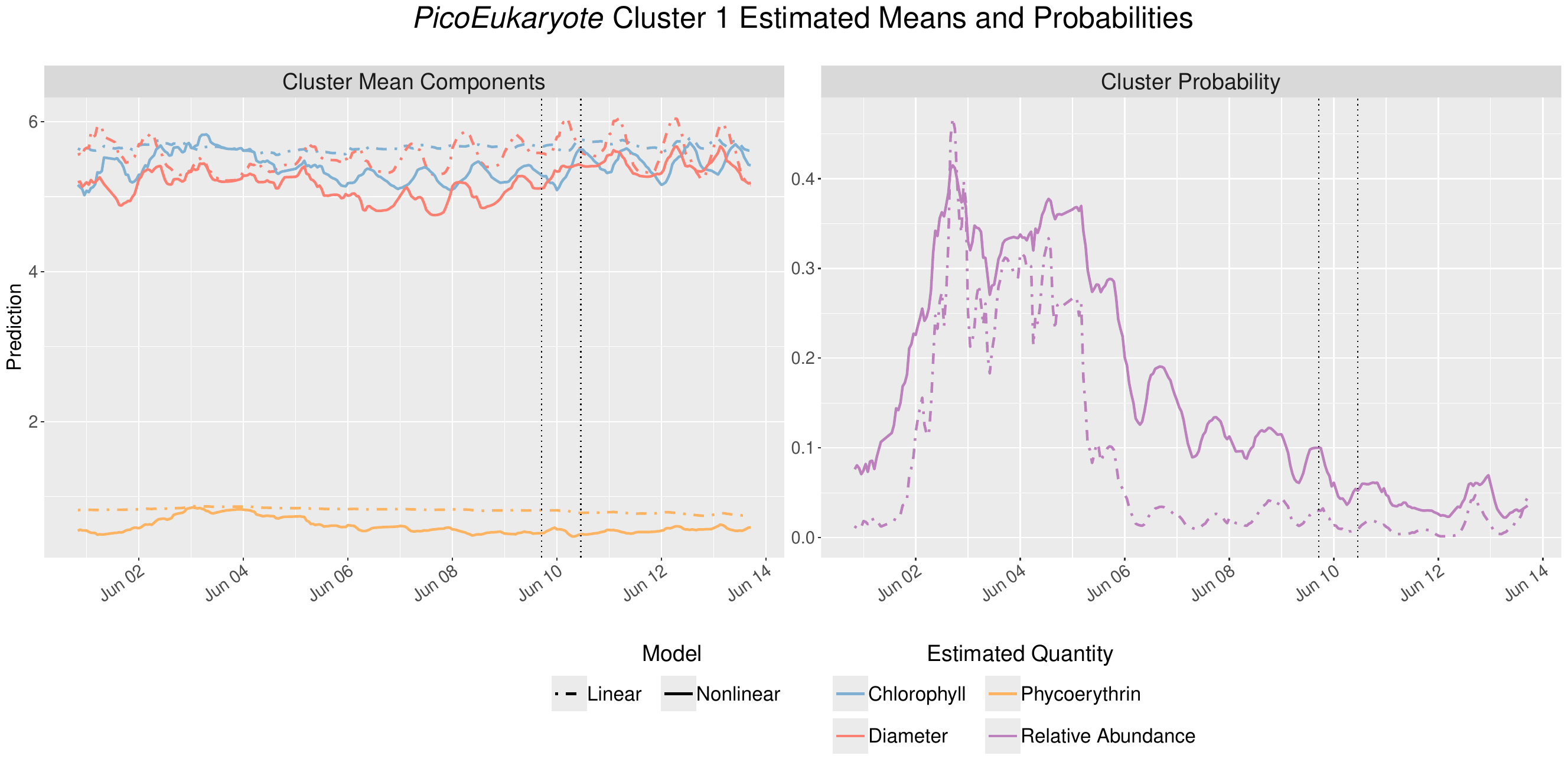}
    \caption{\textit{Estimated means and probabilities over time of one of the \textup{PicoEukaryote} clusters (Pico1) for each model.}}
    \label{fig:Sim_gate_pico1}
\end{figure}

\begin{figure}[H]
    \centering
    \includegraphics[width=\linewidth]{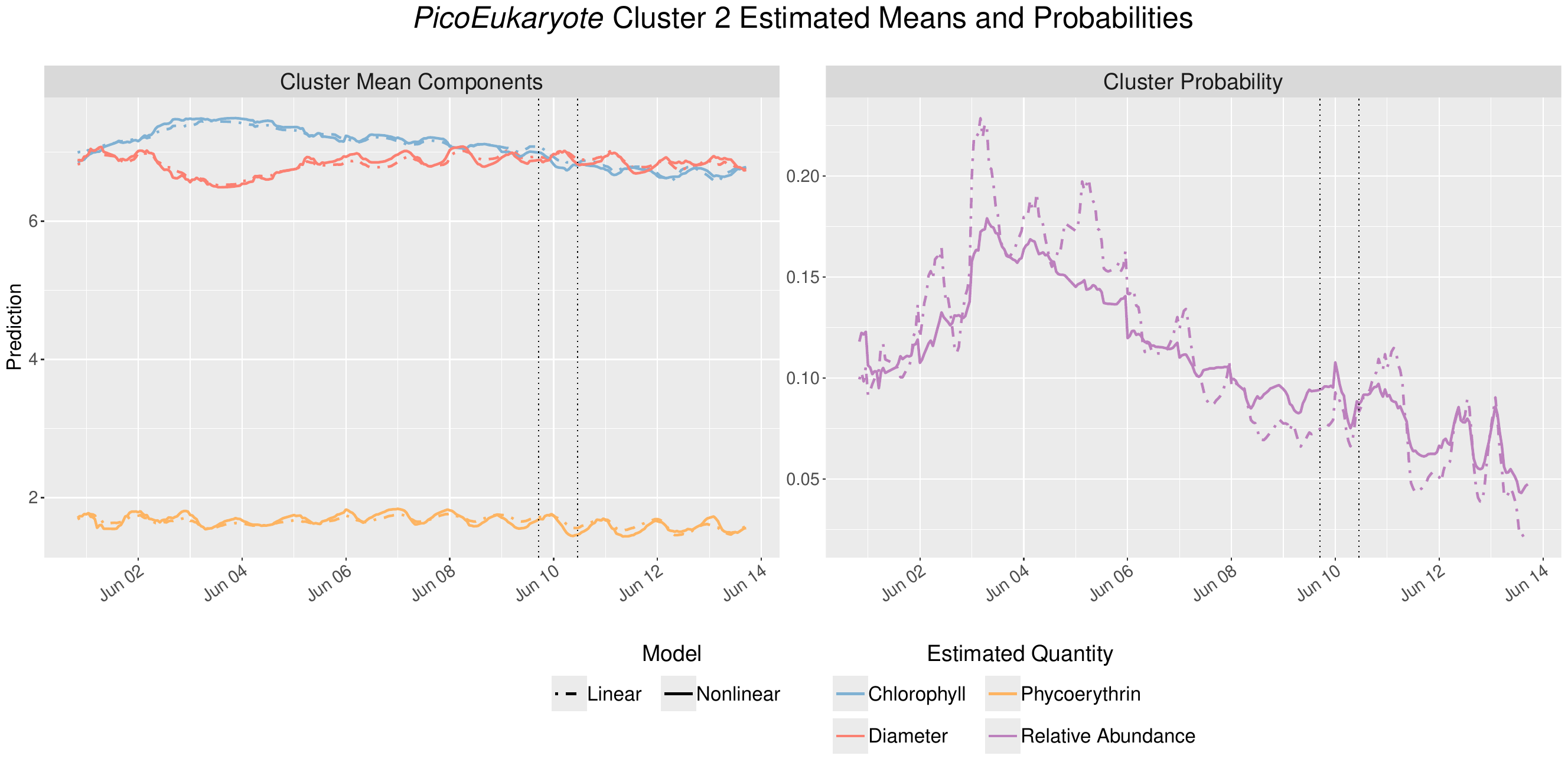}
    \caption{\textit{Estimated means and probabilities over time of one of the \textup{PicoEukaryote} clusters (Pico2) for each model.}}
    \label{fig:Sim_gate_pico2}
\end{figure}

\section*{Supplement C: Partial Response Curves}
\begin{figure}[H]
    \centering
    \includegraphics[width=\linewidth]{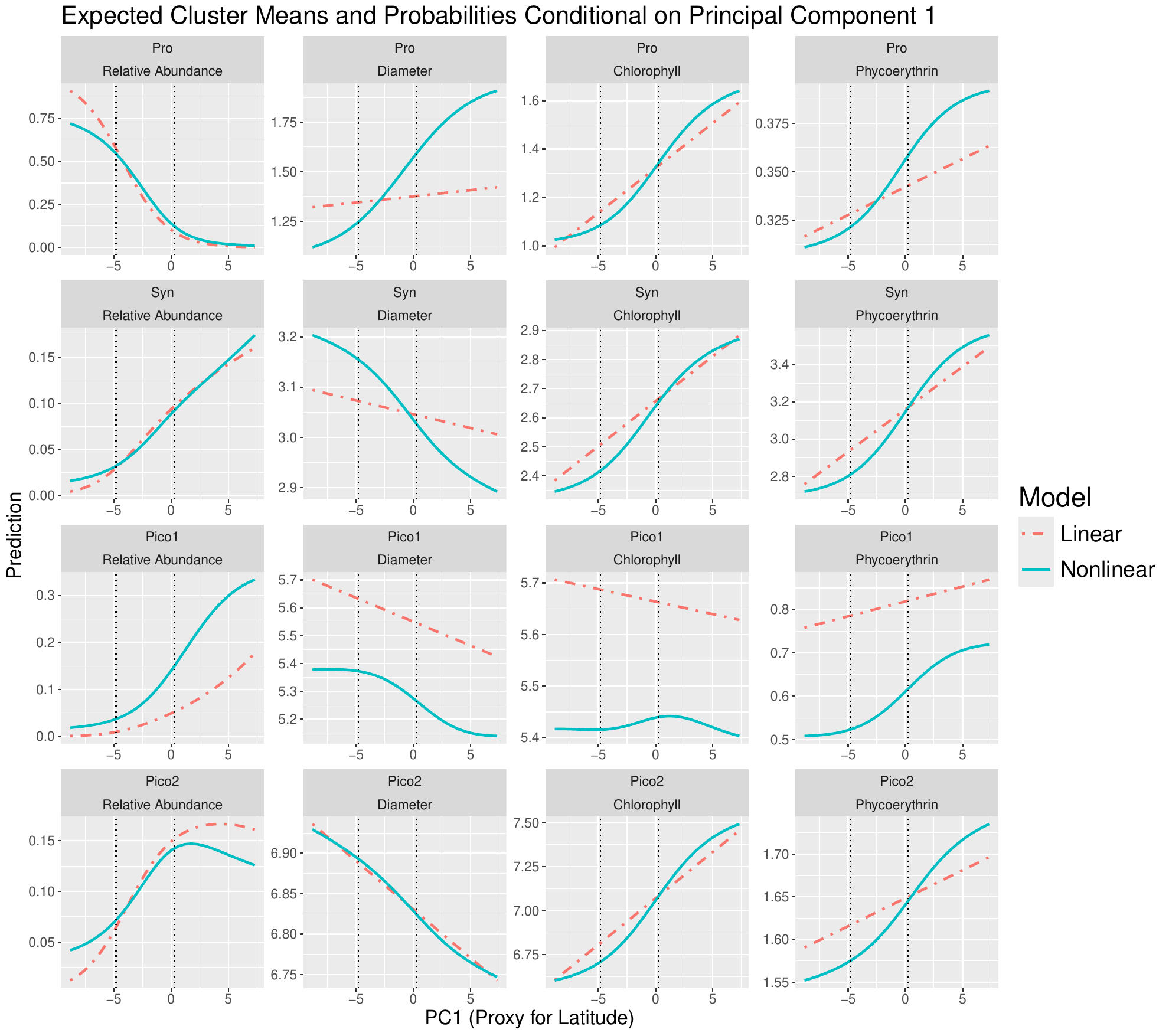}
    \caption{\textit{Partial response curves of each response variable for each species (Pro, Syn, Pico1, and Pico2) as functions of the second principal component.}}
    \label{fig:PC1_PDPs}
\end{figure}

\begin{figure}[H]
    \centering
    \includegraphics[width=\linewidth]{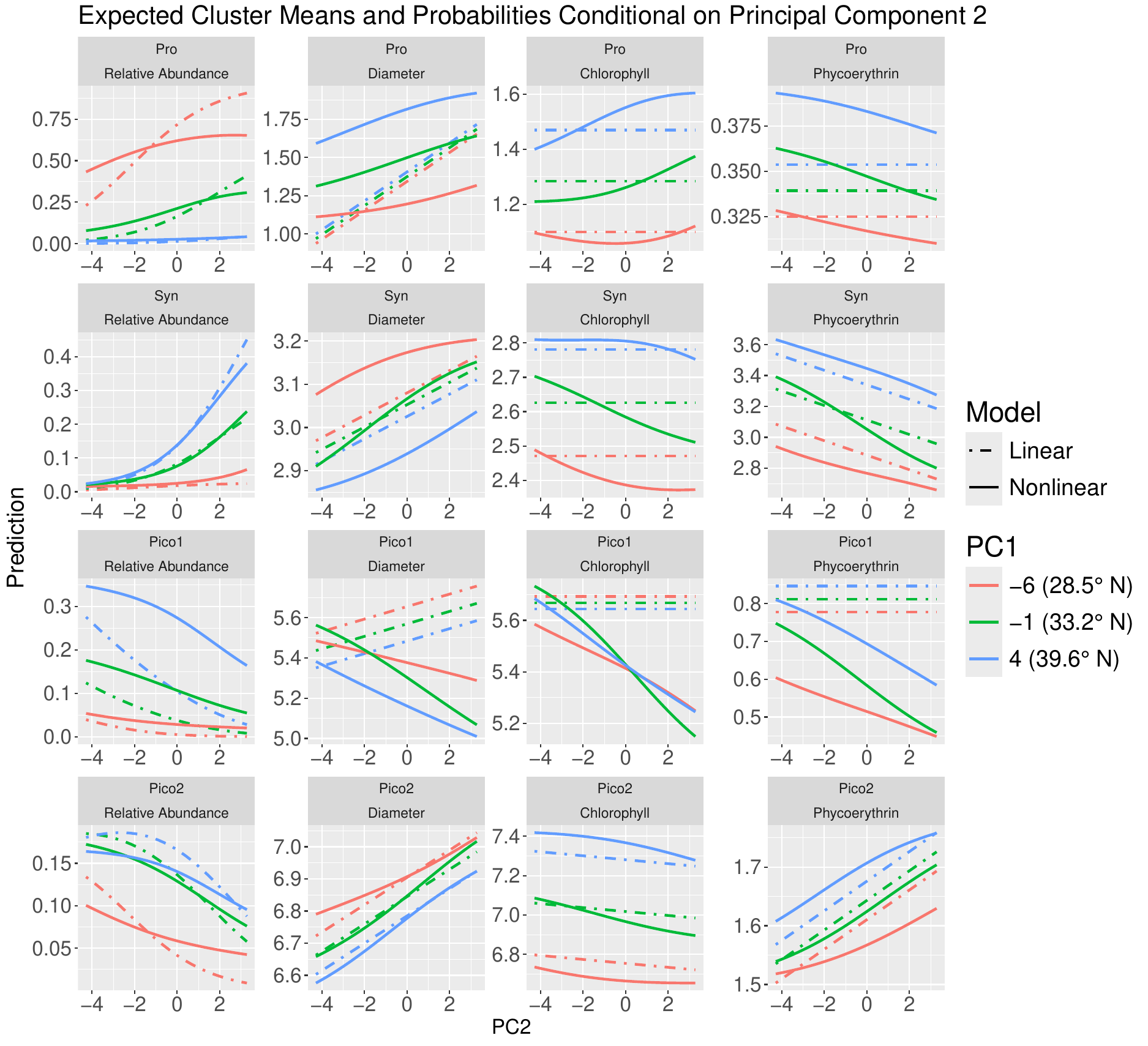}
    \caption{\textit{Partial response curves of each response variable for each species as functions of the second principal component.}}
    \label{fig:PC2_PDPs}
\end{figure}

\section*{Supplement D: Multivariate Simulation and Comparison with Alternative Methods}

We conduct a simple three-dimensional simulation study to demonstrate the ability of the proposed model to recover a mixture of multivariate Gaussian components. For all time points $t$, we simulate $n_t = 1,000$ particles and bin them using $D = 50$ bins in each dimension. The two clusters are generated with the following parameters:
\begin{equation*}
    \begin{aligned}
        \mu_{1,t,1} & = 1.05707 + 0.015 \bPsi_{t, 1} +  0.02 \bPsi_{t, 4} - 0.003 \bPsi_{t, 1} \bPsi_{t, 4}\\
        \mu_{2,t,1} & = 1.463032 \\
        \mu_{1,t,2} = \mu_{2,t,2} & = 1.463032 \\
        \mu_{1,t,3} = \mu_{2,t,3} & = 1.463032 \\
        \text{logit}(\pi_{1,t}) & = - 0.4 \bPsi_{t, 1} + 0.1 \bPsi_{t, 2} \\
        \bs{R}_1 & = \begin{pmatrix}
            1 & 0.5 & 0 \\
            0.5 & 1 & 0 \\ 
            0 & 0 & 1
        \end{pmatrix} \\ 
        \bs{R}_2 & = \begin{pmatrix}
            1 & -0.7 & 0 \\ 
            -0.7 & 1 & 0.3 \\ 
            0 & 0.3 & 1
        \end{pmatrix} \\ 
        \bs{\sigma}_1 = \bs{\sigma}_2 & = (0.2, 0.15, 0.25)',
    \end{aligned}
\end{equation*}
where $\bs{R}_k$ and $\bs{\sigma}_k$ denote the correlation matrix and vector of standard deviations of cluster $k$, respectively, and $\mu_{k,t,\ell}$ denotes the $\ell$th component of the mean of cluster $k$ at time $t$, for $\ell = 1, 2, 3$. The covariates $\bPsi$ are defined in Section 2 of the main text.

\begin{figure}[H]
    \centering
    \includegraphics[width=0.9\linewidth]{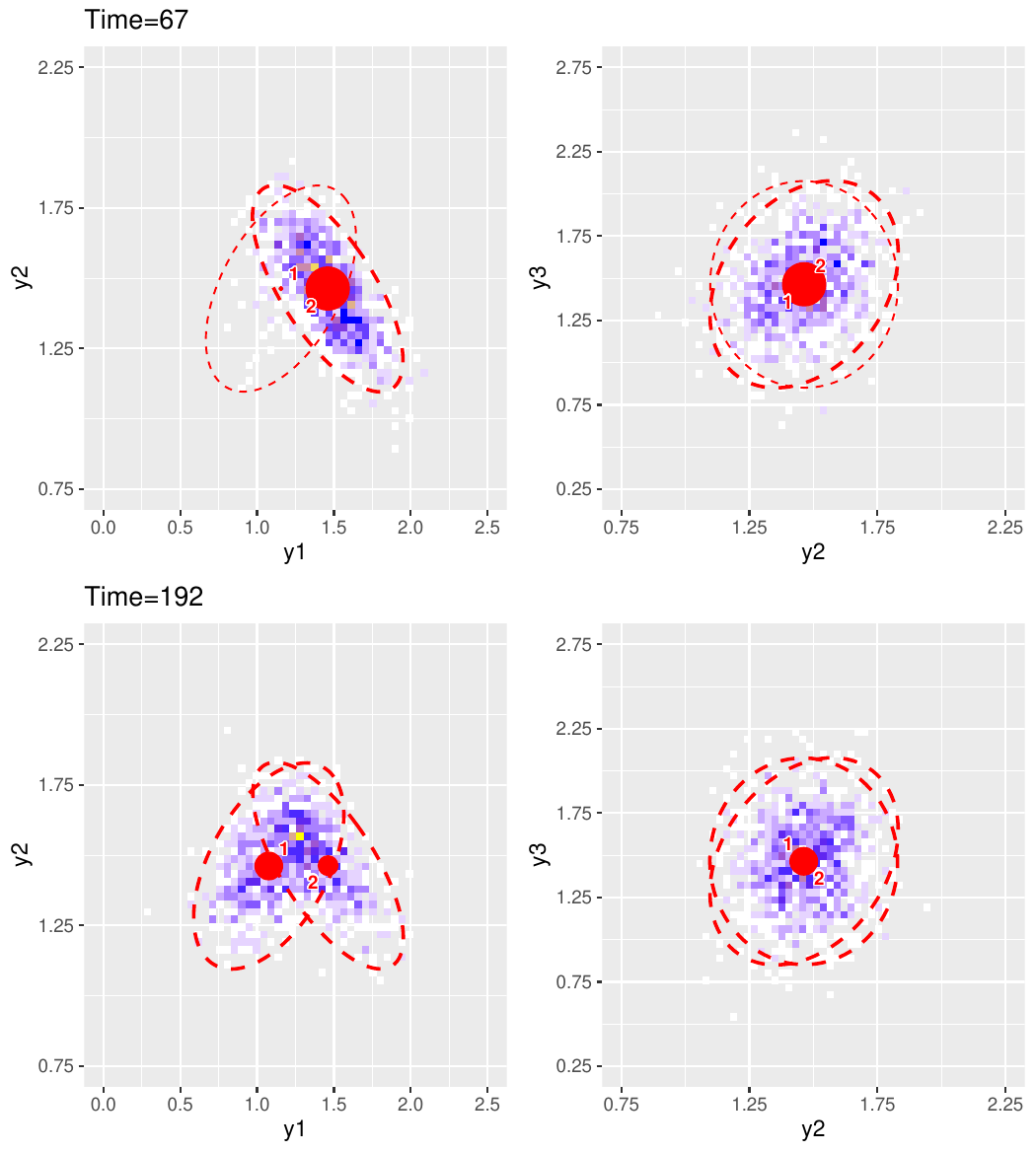}
    \caption{\textit{Three-dimensional simulated cytograms at two time points, $t = 67$ and $t = 192$, with nonlinear model estimates overlaid in red. The data in the background are binned, with the intensity of blue hue proportional to the total number of particles in each bin. The solid points mark the cluster means, and the sizes of the points are proportional to the estimated cluster probabilities. The ellipses represent 2-dimensional projections of symmetric cluster-conditional 95\% probability sets of each cluster.}}
    \label{fig:Fig15_comp}
\end{figure}

\begin{figure}[H]
    \centering
    \includegraphics[width=\linewidth]{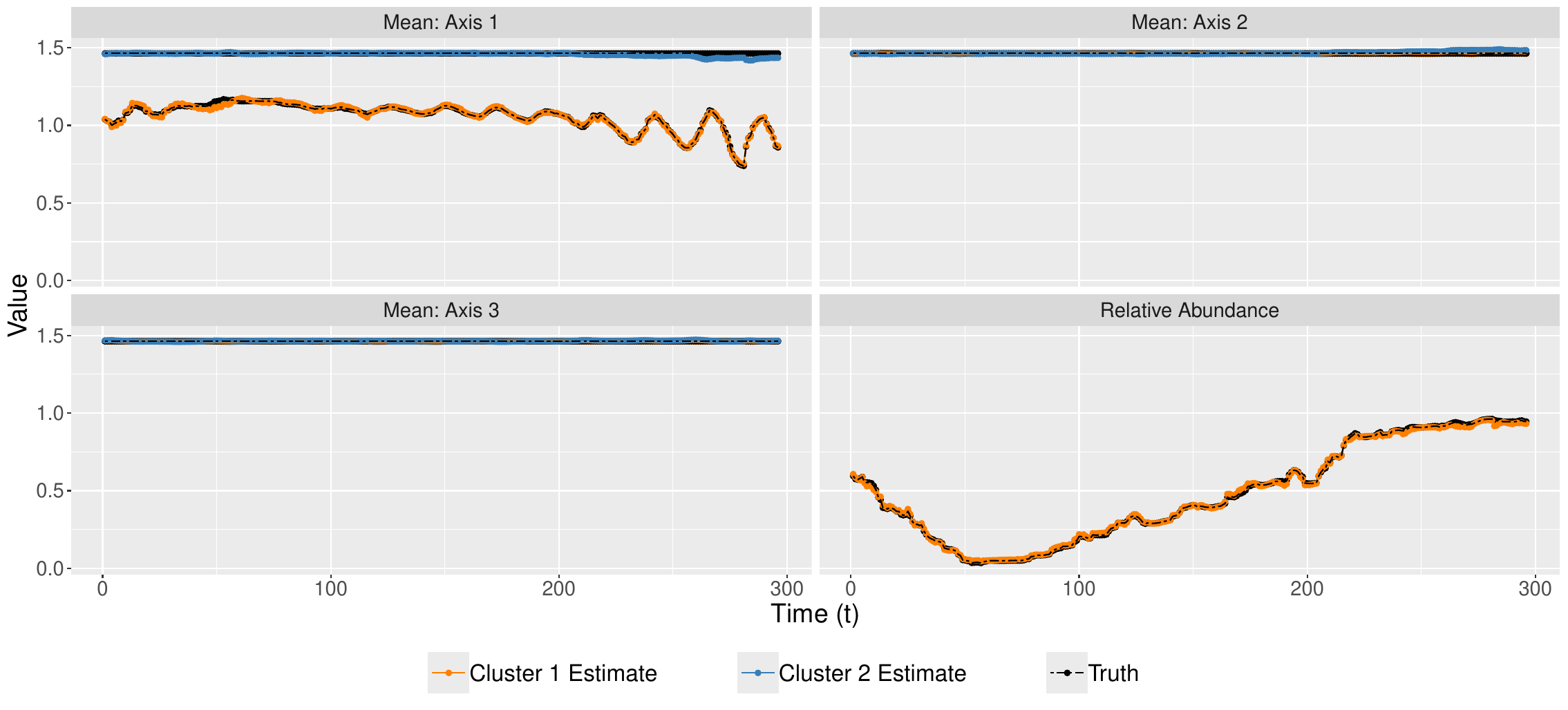}
    \caption{\textit{Mean and probability estimates over time for two three-dimensional simulated clusters, compared with ground truth cluster parameters. The three dimensions of the cluster mean are displayed in the top-left, top-right, and bottom-left panels. The bottom-right panel compares the estimates of the mixing probability of the first cluster with the true probability. The estimated parameters for the first and second clusters are represented by orange and blue solid lines, respectively. True parameter values are represented by black dot-dashed lines.}}
    \label{fig:Fig15_res}
\end{figure}

\autoref{fig:Fig15_comp} overlays the estimated clusters on the data at two representative time points. Although the clusters exhibit substantial overlap at $t = 67$, the proposed model accurately recovers their locations and covariance structure by borrowing information from time points with greater separation, such as $t = 192$. \autoref{fig:Fig15_res} shows that the cluster means and mixing probabilities are recovered with near-perfect accuracy.

Additionally, we fit a variety of existing clustering and regression methods to this dataset to compare their performance against the proposed model. In particular, we consider the method of \textcite{flowmix}, generalized additive models \parencite{Wood2017}, random forests \parencite{Breiman2001}, and the mixture of experts distributional regression method of \textcite{Rugamer2024}, hereafter referred to as MoE-DR. Since the latter three methods were not originally designed for multivariate mixture regression with replicated responses, each required adaptations to make the comparison possible. For instance, they expect a single response observation per covariate observation, whereas the simulated dataset contains multiple $\boldsymbol{y}_i^{(t)}$ values for every $\boldsymbol{x}_t$. Consequently, we replicated the covariates to obtain pairs $\left(\boldsymbol{y}_i^{(t)}, \boldsymbol{x}_i^{(t)}\right)$, where, for every $t = 1, \cdots, T$ and every $i = 1,\cdots, n_t$, $\boldsymbol{x}_i^{(t)} = \boldsymbol{x}_t$. Furthermore, in all methods we assume the true number of clusters is known \textit{a priori}. Unless otherwise mentioned, hyperparameters for each method were selected via 5-fold cross-validation over a grid of 10 candidate values, choosing the value that minimized the negative log-likelihood weighted by particle biomasses. All analyses were conducted in R version 4.4.3 \parencite{R}.

\subsection*{Generalized Additive Model}

Since traditional generalized additive models (GAMs) cannot cluster data, we adopt a three-stage approach: cluster, then align cluster labels across samples, and then regress. Treating clustering, population alignment, and downstream analysis as separate steps is a common workflow in flow cytometry \parencite{Azad2012}, making this comparison representative of existing analysis pipelines. More specifically, we first use $k$-means clustering to assign cluster memberships to each observation within each cytogram. For each cytogram, we repeat the clustering with 10 different initializations and retain the clustering which minimizes the ratio of the within-cluster sum of squares to the total sum of squares. Second, in order to correct for label switching between cytograms, we keep the cluster labels in cytogram 1 fixed. Then, for $t$ in $\{2, \cdots, T\}$, we reassign the cluster labels in cytogram $t$ using the permutation of labels which minimizes the Kullback-Leibler divergence from cytogram $t - 1$ to cytogram $t$. In the third stage, we treat the cluster labels as known and fit a separate GAM to each cluster. In particular, we use the \texttt{bam} function \parencite{Wood2015} in the \texttt{mgcv} package for R \parencite{Wood2011} with penalized cubic regression splines (\texttt{cs}). After replicating the covariates as described above, the size of the dataset was too large for this package's implementation of multivariate regression, so we fit a separate GAM to each response dimension. As a result, estimates of each cluster's covariance matrix are restricted to be diagonal. Additionally, we passed the biomasses $c_i^{(t)}$ to be used as model weights. We allowed smooth effect selection and selected the effective degrees of freedom scaling factor via cross-validation. Finally, parameters were estimated via fast restricted maximum likelihood estimation.

\subsection*{Random Forest}

Like GAMs, random forests do not directly model latent mixture components, so we adopt the same three-stage cluster-align-regress framework. Instead of $k$-means, we use the \texttt{sidClustering} algorithm \parencite{Mantero2021} as implemented in the \texttt{randomForestSRC} R package \parencite{Ishwaran2026}. Due to the large size of the dataset, we used fast approximate random forests (\texttt{rfsrc.fast}) in the regression stage. To account for the class imbalance, we used the biomass weights $c_i^{(t)}$ as bootstrap sampling weights, thereby increasing the probability of sampling large but rare individuals. All other settings were left at their default settings: 500 trees, $p/3$ covariates used as candidate variables at each split for regression and $\lceil\sqrt{p}\rceil$ for classification, 10 random split points per variable, a minimum terminal node size of 5 in regression and 1 in classification, and a bootstrap sample size of $\lceil0.632N\rceil$. Additionally, we disabled variable selection during clustering.

\subsection*{Mixture of Experts Distributional Regression}

MoE-DR assumes univariate responses, so we first tried fitting individual models to each response dimension as with the GAM experiment. However, we found that since the mixture of experts framework jointly estimates cluster memberships and regression parameters, the three models may not agree on each particle's cluster membership. To remedy this, we chose to transform the response data using principal components analysis. We assumed the first principal component approximately captured the primary direction separating the mixture components. Accordingly, we fit the mixture of experts to the first principal component using the \texttt{mixdistreg} R package \parencite{mixdistreg}, then fit a single-component distributional regression model to each of the remaining principal components. Further, for each principal component, we fit $K$ distributional regressions. For $k$ in $\{1, \cdots, K\}$, we fit a distributional regression where the likelihood contribution of each particle is reweighted by its posterior probability of membership in cluster $k$. That is, we rescale the original biomass weight $c_i^{(t)}$ by multiplying it by the observation's \textit{responsibility} for component $k$. Rather than assigning cluster memberships and ignoring their uncertainty, this simply allows the parameter estimates of each cluster to be more heavily influenced by observations likely to come from that cluster. This procedure is analogous to a single iteration of an expectation-maximization algorithm. 

For the regression structures, we use neural networks with 2 hidden layers and 64 nodes per layer as the gating network and each expert. We use $\ell_1$ regularization with the penalties for each network (gating and expert) selected via cross-validation. Training was done using the Adam optimizer \parencite{Adam}, a batch size of 512, a patience of 10, and a maximum of 500 epochs. The model parameters yielding the minimum validation loss were retained.

\subsection*{Results}

\begin{figure}
    \centering
    \includegraphics[width=\linewidth]{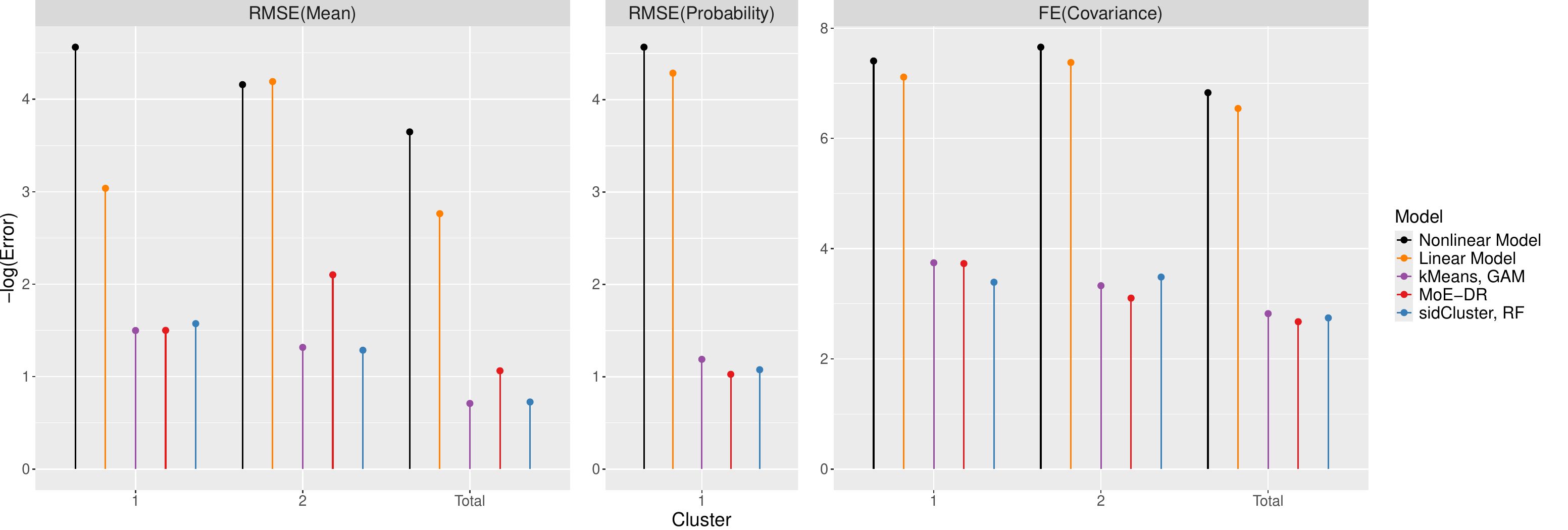}
    \caption{Plot of accuracy metrics of each model by cluster, as well as totals across clusters. The negative logarithm of each error metric is displayed for ease of visualization; larger values indicate more accurate parameter estimates.}
    \label{fig:comparison}
\end{figure}

To compare the models, we evaluated estimation accuracy separately for the cluster means, mixing probabilities, and covariance matrices. In particular, we consider the root mean squared ($\ell_2$) error of the mean
\begin{equation*}
    \mathrm{RMSE}(\hat{\boldsymbol{\mu}}_{k, t} \mid \mid \boldsymbol{\mu}_{k, t}) = \sqrt{\frac{1}{T} \sum_{t = 1}^T \left|\left| \hat{\boldsymbol{\mu}}_{k, t} - \boldsymbol{\mu}_{k, t} \right|\right|^2},
\end{equation*}
and the root mean squared error of the mixing probability 
\begin{equation*}
    \mathrm{RMSE}(\hat{\pi}_{k, t} \mid \mid \pi_{k, t}) = \sqrt{\frac{1}{T} \sum_{t = 1}^T \left( \hat{\pi}_{k, t} - \pi_{k, t} \right)^2},
\end{equation*}
and the Frobenius error of the covariance matrix, which is constant across time:
\begin{equation*}
    \mathrm{FE}\left(\hat{\boldsymbol{\Sigma}}_k \mid \mid \boldsymbol{\Sigma}_k\right) = \left|\left| \hat{\boldsymbol{\Sigma}}_k - \boldsymbol{\Sigma}_k \right|\right|_\mathrm{F}.
\end{equation*}
In \autoref{fig:comparison}, we display the cluster-specific error metrics as well as their total across the two clusters. For ease of visualization, we display the negative logarithm of each metric, so larger values indicate more accurate parameter estimates. Overall, our proposed method (``Nonlinear Model'') and the method of \textcite{flowmix} (``Linear Model'') are much more accurate than the GAM, random forest, and MoE-DR. In the GAM and random forest approaches, the initial hard clustering step may introduce errors by effectively removing observations from certain regions of each component's support, distorting shape and location. Meanwhile, the MoE-DR approach is limited by its inability to model the three response dimensions jointly, requiring strong assumptions to resolve cluster membership. Because GAM and MoE-DR fit each response dimension independently, they cannot estimate correlations between response dimensions, resulting in diagonal covariance estimates. In contrast, the random forest is non-probabilistic and therefore estimates the covariance matrix from regression residuals. Consequently, any error in the fitted mean function propagates to the covariance estimate. 

The linear model is slightly more accurate in estimating the cluster 2 mean, which is constant with respect to the covariates. Since the true regression function for this component is linear, the additional flexibility of the nonlinear model offers little advantage in this setting and may introduce additional estimation variability. Overall, these results suggest that our proposed method more accurately recovers the distributions of latent subpopulations than the competing methods considered here, particularly in covariance estimation and in recovering nonlinear dynamics.

\section*{Supplement E: Robustness to Hidden Layer Width and Randomness}

\begin{figure}[H]
    \centering
    \includegraphics[width=\linewidth]{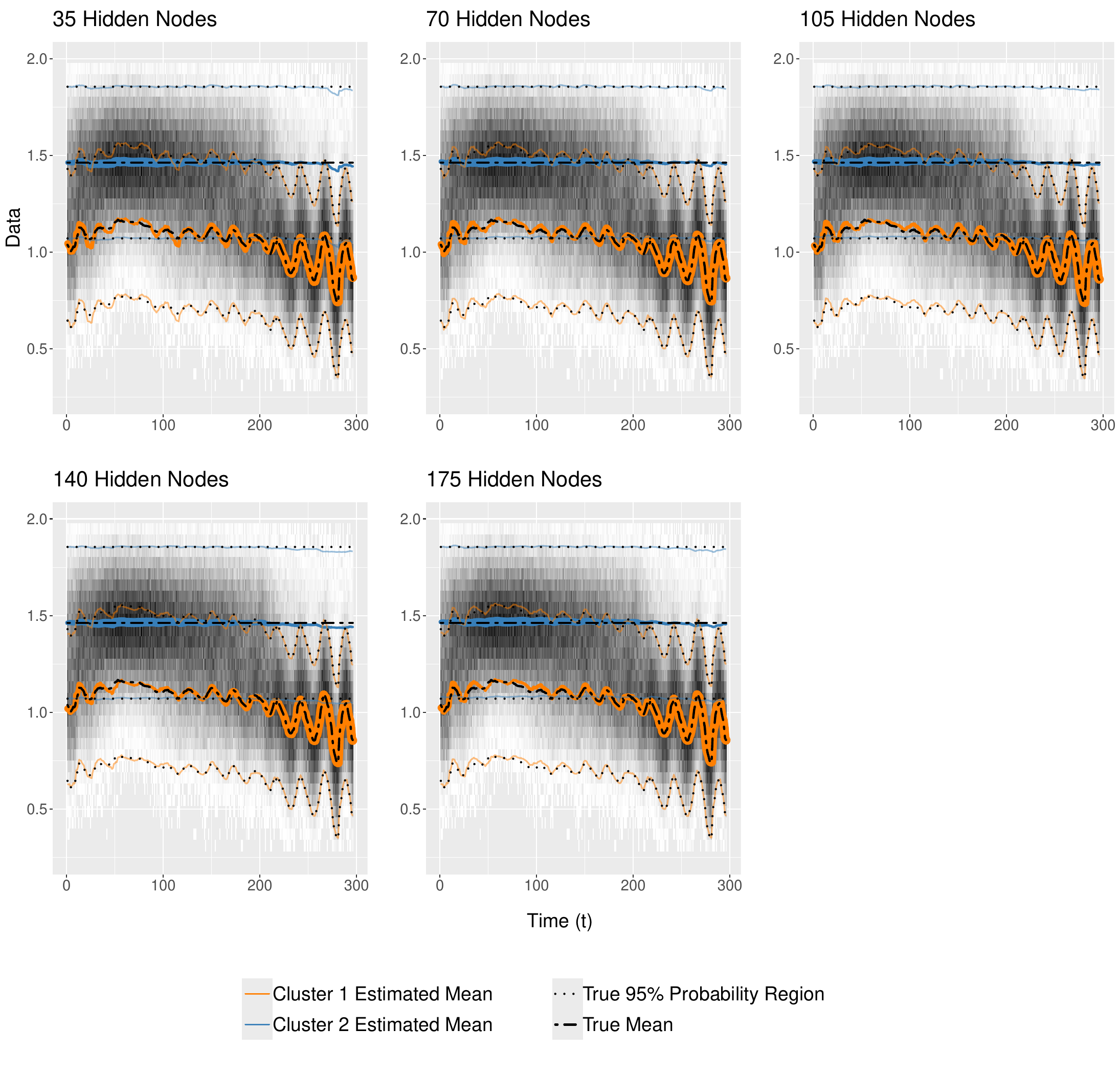}
    \caption{\textit{Estimated nonlinear models on simulated 1-dimensional data with a variety of hidden layer widths. All panels display in the background the simulated and binned data $y_b^{(t)}$ over time $t$, generated from the ``Interaction Mean'' model, with signal size $\Delta = 0.40384$. The greyscale color of the bin $b$ at time $t$ is proportional to the biomass $c_b^{(t)}$. Overlaid are the estimated parameters $\hat{\boldsymbol{\mu}}_{k, t}$ over time $t$ from the nonlinear model as solid colored lines. These lines' thickness at time $t$ is proportional to the estimated cluster probability $\hat \pi_{k,t}$. The thin colored lines are the bounds of symmetric 95\% pointwise conditional probability regions of each Gaussian.
    The black dashed lines represent the true cluster means, and the black dotted lines represent the true 95\% central probability regions.}}
    \label{fig:widths_experiment}
\end{figure}

To assess the robustness of our approach to the choice of hidden layer width (number of hidden nodes), we repeat the experiment shown in Figure 2 of the main text, for hidden widths $n_h = 35, 70, 105, 140$, and 175. As shown in \autoref{fig:widths_experiment}, the resulting estimates are nearly indistinguishable across these settings. However, the estimated mean of cluster 2 deviates from the true mean as the hidden width moves farther from $n_h = 70$. This choice of $n_h = 70$ for the main results in this article therefore represents a reasonable compromise between flexibility and stability in estimating nonlinear functions. 

\begin{figure}[H]
    \centering
    \includegraphics[width=\linewidth]{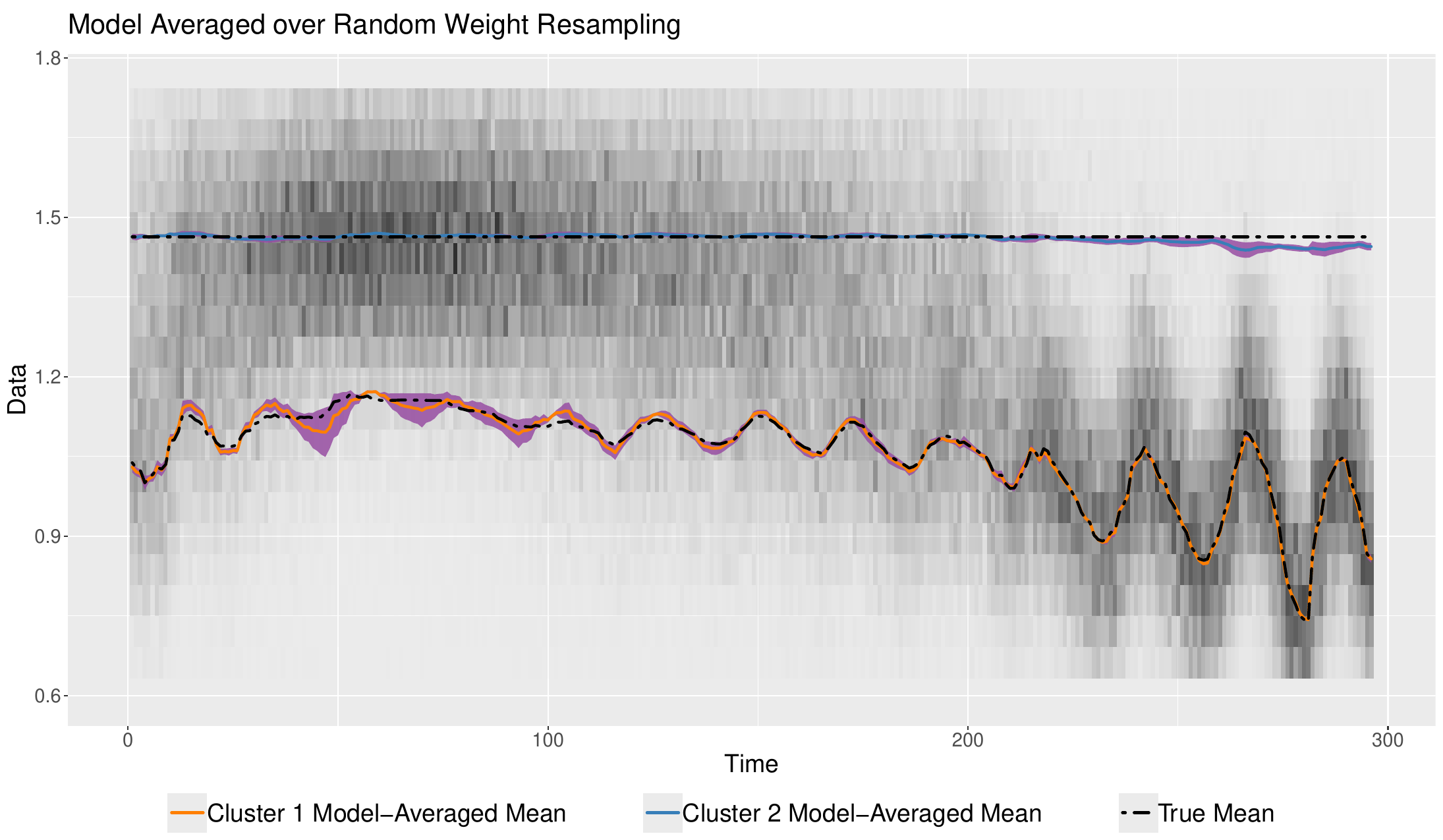}
    \caption{\textit{Average of 5 estimated nonlinear models with randomly resampled hidden layer weights on simulated 1-dimensional data. All panels display in the background the simulated and binned data $y_b^{(t)}$ over time $t$, generated from the ``Interaction Mean'' model, with signal size $\Delta = 0.40384$. The greyscale color of the bin $b$ at time $t$ is proportional to the biomass $c_b^{(t)}$. These lines' thickness at time $t$ is constant with respect to $\hat \pi_{k,t}$ in order to avoid obscuring the uncertainty intervals. Approximate 95\% confidence intervals around the model average---assuming a $t$-distribution with 4 degrees of freedom---are highlighted in purple.}}
    \label{fig:seeds_experiment}
\end{figure}

Resampling the hidden weights may introduce small variations in the estimates, but the method is generally robust to this randomness. In \autoref{fig:seeds_experiment} we display mean estimates averaged over five models with $n_h = 70$ hidden widths, each fit using weights generated from different random number generator seeds. Overall, the estimates are very similar across models. Some variability is observed during the interval $t \in [10, 150]$, where the amplitude of the oscillating mean changes. The model's ability to capture nonlinear behavior therefore depends to some extent upon the particular draw of the hidden layer weights.

To increase the chances of obtaining favorable weights, two strategies may be considered. One could estimate multiple models with different random draws of the hidden weights and select the model with the lowest out-of-sample negative log-pseudolikelihood. Alternatively, one could estimate a single model with a large number of hidden nodes and allow the variable selection property of the lasso penalties to retain the hidden weights that contribute most to predictive performance. We prefer the latter approach, since it reduces the number of hyperparameters to tune and is less computationally demanding, requiring estimation of only a single model.

\section*{Supplement F: Number of Principal Components}

\begin{figure}[H]
    \centering
    \includegraphics[width=\linewidth]{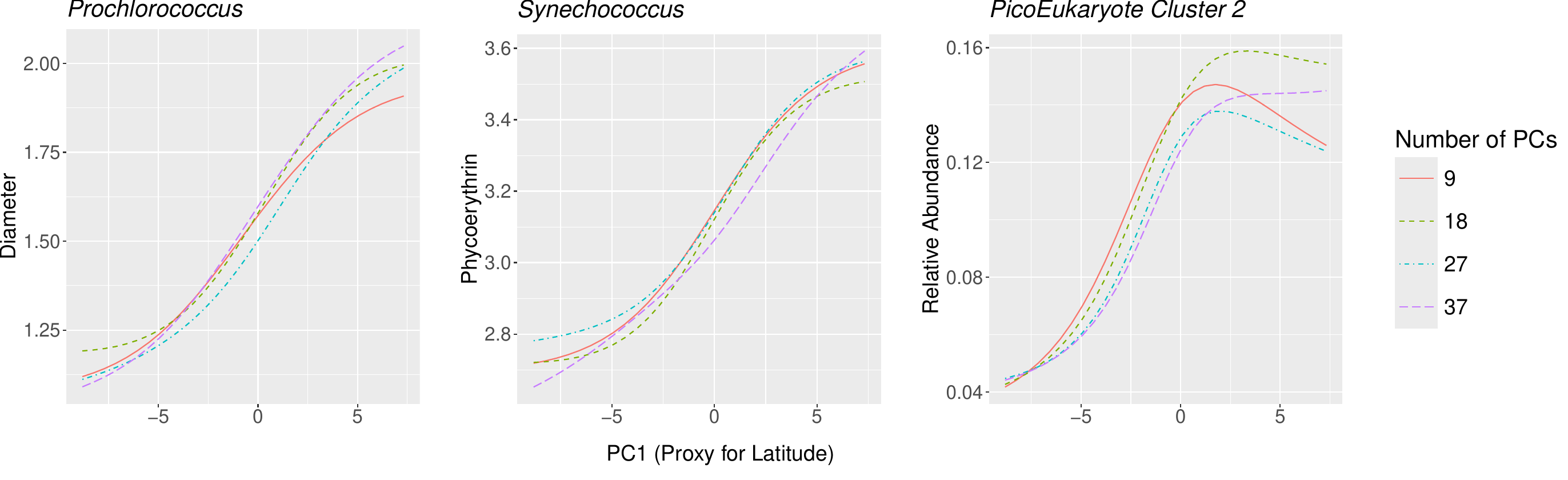}
    \caption{\textit{Partial responses of \textup{Prochlorococcus} diameter, \textup{Synechococcus} phycoerythrin fluorescence, and the second \textup{PicoEukaryote} cluster's relative abundance as functions of the first principal component. A separate curve is plotted for each number of principal components used as regressors in the nonlinear model. Predictions from the models using 9, 18, 27, and 37 principal components are represented by the solid, short-dashed, dot-dashed, and long-dashed lines, respectively.}}
    \label{fig:n_PCs}
\end{figure}

Here, we evaluate the stability of our model to the number of principal components used as regressors. We repeat the experiment in Section 5 of the main text, with $q = 9, 18, 27$, and $37$ principal components and with the number of hidden nodes fixed at $n_h = 70$. Since the principal components are used to improve interpretation, rather than model fit, we evaluate stability based on the variability of the partial response curves over different values of $q$, as displayed in \autoref{fig:n_PCs}. There are some discrepancies in the estimated \textit{PicoEukaryote} cluster 2 relative abundances at large values of PC1, which shows the potential usefulness of the trailing principal components beyond the first handful. But overall, the partial response curves change very little as a function of $q$. Additionally, the average 5-fold cross-validation score across the four models was 3.240 with a standard error of 0.016, indicating nearly identical predictive performance.

\section*{Supplement G: Effect of Likelihood Biomass Weights}

In order to avoid underestimating the prevalence of massive but rare particles, we raise the density of each particle $\bs{y}_b^{(t)}$ to a power proportional to the total biomass within that bin, $\bs{c}_b^{(t)}$, as shown in Equation (3) in the main text. Here, we compare this approach to the \textit{binned counts} representation of the data, in which the power is simply the number of particles within the bin. The likelihood resulting from the binned counts representation is proportional to the likelihood of a discretized sample of the data with bin $b$ observed $c_b^{(t)}$ times. It is therefore a proper likelihood.

To showcase the effect of using  biomass weights on parameter estimation, we cross-validate and refit our proposed model on the SeaFlow cruise data as in Section 5 of the main text twice: once with the binned biomass representation, and again with the binned counts representation. 

\begin{figure}[H]
    \centering
    \includegraphics[width=\linewidth]{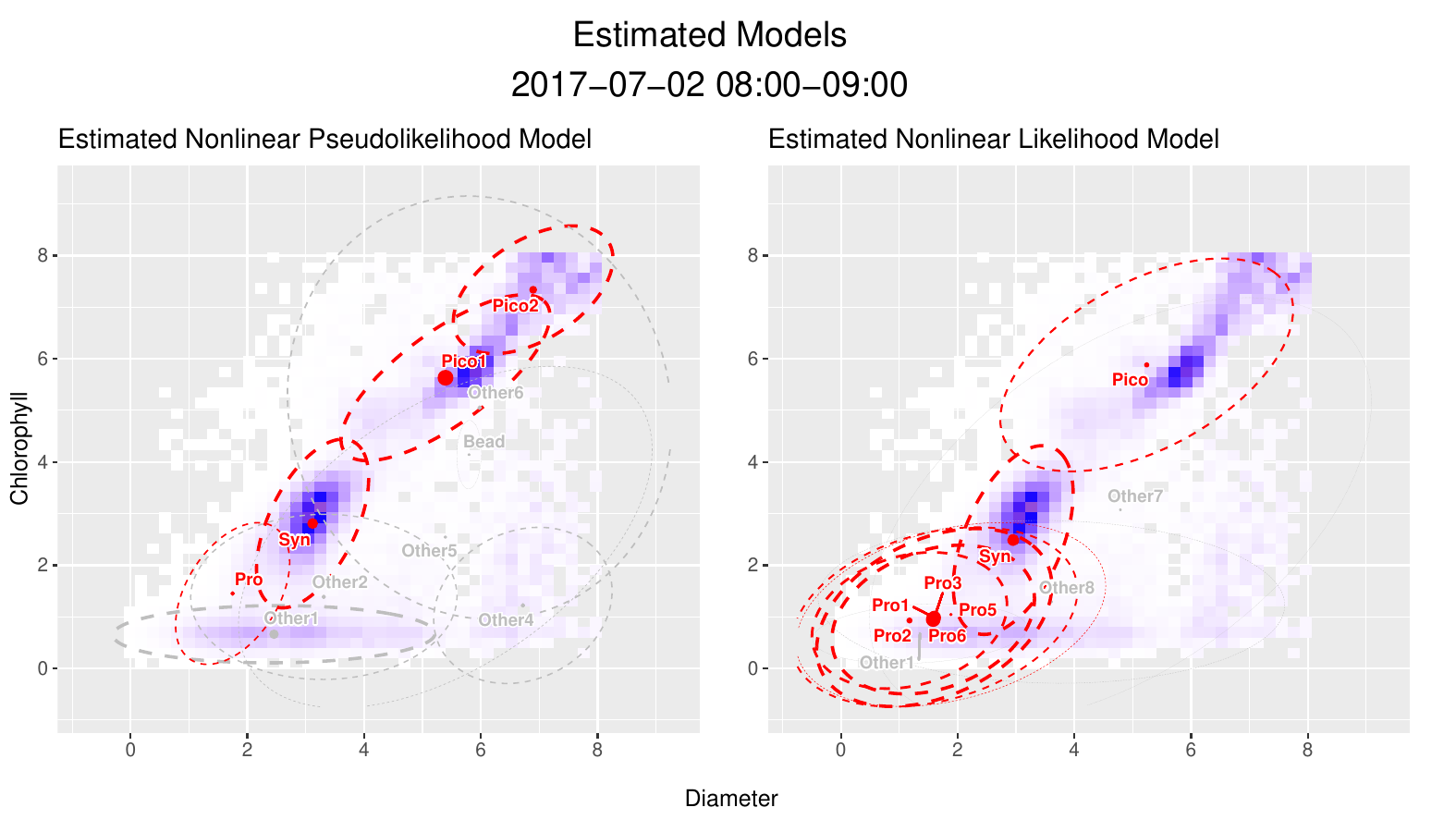}
    \caption{\textit{Cytogram in two dimensions, diameter and chlorophyll, at time $t = 33$ (2017-07-02 08:00-09:00), with model estimates overlaid in red (marking the important clusters) or grey (the rest). The data in the background are binned, with the intensity of blue hue proportional to the total biomass in each bin. The solid points mark the cluster means, and the sizes of the points are proportional to the estimated cluster probabilities. The ellipses represent 2-dimensional projections of symmetric cluster-conditional 95\% probability sets of each cluster. The model estimates on the left are obtained using scaled particle biomasses as likelihood weights, while the estimates on the right are obtained using weights proportional to the number of particles within each bin.}}
    \label{fig:wts-comp}
\end{figure}

Expert annotation by manual gating shows that there are four photosynthetic clusters: \textit{Prochlorococcus}, \textit{Synechococcus}, and \textit{PicoEukaryote} 1 and 2. 
As we see in the left panel of \autoref{fig:wts-comp} using a pseudolikelihood (bins being upweighted by their biomass) recovers these expert-annotated clusters much more precisely, with each 95\% high density region tightly enclosing high-biomass regions of cytogram space and reasonable probability estimates based on the amount of biomass seen within each cluster. 

On the other hand, the right panel of \autoref{fig:wts-comp} clearly shows that omitting the biomass weights leads the model estimate to have many clusters whose cell diameter and chlorophyll are small. There are 6 estimated clusters tightly bunched around the general region where \textit{Prochlorococcus} is located, with many of them highly overlapping. Only one potential large Gaussian cluster captures where \textit{PicoEukaryotes} generally are. This cluster, as well as \textit{Synechococcus}, have means and 95\% high density regions which do not concentrate precisely around high biomass regions. This shows that the likelihood-based model which is based on the relative \textit{counts} of the particles, is inferior in its ability to cluster populations compared to the pseudolikelihood-based model that is based on the relative \textit{biomass contribution} of each subpopulation -- the latter being more scientifically relevant.

\newpage
\printbibliography
    
\end{refsection}

\end{document}